\documentclass[dvips,12pt,a4paper]{article}
\usepackage{a4wide}
\usepackage{epsfig}
\usepackage{amsmath}
\usepackage{latexsym}

  \newlength{\absize}
\newcommand{\half}{{\textstyle\frac{1}{2}}}

\def\lsim{\mathrel{\rlap{\raise 2.5pt \hbox{$<$}}\lower 2.5pt \hbox{$\sim$}}}

\renewcommand{\Re}{\mbox{Re}}

\newcommand{\Lumint}{{\cal L}_{\rm int}}

\def\ta{\tilde{a}}
\def\tb{\tilde{b}}
\def\tc{\tilde{c}}
\def\td{\tilde{d}}
\def\te{\tilde{e}}

\renewcommand{\theequation}{\thesection.\arabic{equation}}

\allowdisplaybreaks 

\catcode`@=11
\def\citer{\@ifnextchar [{\@tempswatrue\@citexr}{\@tempswafalse\@citexr[]}}
 
\def\@citexr[#1]#2{\if@filesw\immediate\write\@auxout{\string\citation{#2}}\fi
  \def\@citea{}\@cite{\@for\@citeb:=#2\do
    {\@citea\def\@citea{--\penalty\@m}\@ifundefined
       {b@\@citeb}{{\bf ?}\@warning
       {Citation `\@citeb' on page \thepage \space undefined}}%
\hbox{\csname b@\@citeb\endcsname}}}{#1}}
\catcode`@=12

\begin{document}
  \thispagestyle{empty}
  \pagestyle{empty}
  \renewcommand{\thefootnote}{\fnsymbol{footnote}}
\newpage\normalsize
    \pagestyle{plain}
    \setlength{\baselineskip}{4ex}\par
    \setcounter{footnote}{0}
    \renewcommand{\thefootnote}{\arabic{footnote}}
\newcommand{\preprint}[1]{%
  \begin{flushright}
    \setlength{\baselineskip}{3ex} #1
  \end{flushright}}
\renewcommand{\title}[1]{%
  \begin{center}
    \LARGE #1
  \end{center}\par}
\renewcommand{\author}[1]{%
  \vspace{2ex}
  {\Large
   \begin{center}
     \setlength{\baselineskip}{3ex} #1 \par
   \end{center}}}
\renewcommand{\thanks}[1]{\footnote{#1}}
\renewcommand{\abstract}[1]{%
  \vspace{2ex}
  \normalsize
  \begin{center}
    \centerline{\bf Abstract}\par
    \vspace{2ex}
    \parbox{\absize}{#1\setlength{\baselineskip}{2.5ex}\par}
  \end{center}}

\hyphenation{phenomeno-logy}
\renewcommand{\thefootnote}{\fnsymbol{footnote}}
\begin{flushright}
{\setlength{\baselineskip}{2ex}\par
{March 2004}           \\
} 
\end{flushright}
\vspace*{4mm}
\vfill
\title{Graviton-induced Bremsstrahlung at $e^+e^-$ colliders}
\vfill
\author{
Trygve Buanes\footnote{\tt trygve.buanes@ift.uib.no},
Erik W. Dvergsnes\footnote{\tt erik.dvergsnes@ift.uib.no},
Per Osland\footnote{\tt per.osland@ift.uib.no}}
\begin{center}
Department of Physics and Technology, University of Bergen, \\ All\'{e}gaten
55, N-5007 Bergen, Norway
\end{center}
\vfill

\abstract{We consider graviton-induced Bremsstrahlung at future $e^+e^-$
colliders in both the ADD and RS models, with emphasis on the photon
perpendicular momentum and angular distribution. The photon spectrum is shown
to be harder than in the Standard Model, and there is an enhancement for
photons making large angles with respect to the beam. In the ADD scenario, the
excess at large photon perpendicular momenta should be measurable for values
of the cut-off up to about twice times the c.m.\ energy. In the RS scenario,
radiative return to graviton resonances below the c.m.\ energy can lead to
large enhancements of the cross section.}
\vspace*{20mm} 
\setcounter{footnote}{0} 
\vfill

\newpage
\setcounter{footnote}{0}
\renewcommand{\thefootnote}{\arabic{footnote}}

\section{Introduction}
Early ideas on brane world scenarios date back to more than 15 years ago
\cite{Akama:jy,Antoniadis:1990ew}.  In recent years, more predictive and
explicit scenarios involving extra dimensions have been proposed
\citer{Arkani-Hamed:1998rs,Randall:1999vf}. As opposed to string theory with
tiny compactification scales of ${\cal O}(10^{-35}$~m), there is now a large
number of theories which actually will be tested in the current and next
generation of experiments.

Here we shall consider two of these scenarios, namely the
Arkani-Hamed--Dimopoulos--Dvali (ADD) \cite{Arkani-Hamed:1998rs} and the
Randall--Sundrum (RS) scenario \cite{Randall:1999ee}, and investigate some
signals characteristic of such models at possible future electron--positron
linear colliders like TESLA \cite{Aguilar-Saavedra:2001rg} and CLIC
\cite{Assmann:2000hg}.

The most characteristic feature of these models is that they predict the
existence of massive gravitons, which may either be emitted into the final
state (leading to events with missing energy and momentum), or exchanged as
virtual, intermediate states.  We shall here focus on the effects of such
massive graviton exchange on the Bremsstrahlung process:
\begin{equation}
\label{Eq:ee-mumuga}
e^+e^-\to \mu^+\mu^-\gamma,
\end{equation}
for which the basic electroweak contributions are well known
\cite{Berends:1980yz}.

Due to an extra photon in the final state, this process has a reduced cross
section as compared to two-body final states like $\mu^+\mu^-$ and
$\gamma\gamma$, and is unlikely to be the discovery channel, but it may
provide additional confirmation if a signal should be observed in the two-body
final states. In particular, the presence of additional Feynman diagrams,
without the infrared and collinear singularities of the Standard Model (SM)
leads one to expect a harder photon spectrum.

We shall first, in Sect.~\ref{sec:gen-brems}, present the differential cross
section for the process (\ref{Eq:ee-mumuga}). Integrated cross sections as
well as photon perpendicular momentum and angular distributions will be
discussed. Then, in Sects.~\ref{sec:ADD} and \ref{sec:RS}, we specialize to
the ADD and RS scenarios, by performing sums over the respective KK towers.
In Sect.~\ref{sec:concl} we summarize our conclusions.

\section{Graviton induced Bremsstrahlung}
\label{sec:gen-brems}
In this section we present the cross section for the process
(\ref{Eq:ee-mumuga}), taking into account the $s$-channel exchange of the
photon, the $Z$ and a single graviton of mass $m_{\vec n}$ and width
$\Gamma_{\vec n}$. These results are for the differential cross section very
similar to those obtained for graviton exchange in the analogous process
$q\bar q\to e^+e^-\gamma$ \cite{Dvergsnes:2002nc}, and will in
Sects.~~\ref{sec:ADD} and \ref{sec:RS} be adapted to the ADD and RS scenarios.
\subsection{Differential cross sections} \label{sec:diff-sigma}
The cross section for the process in Eq.~(\ref{Eq:ee-mumuga}) is determined by
the Feynman diagrams of Fig.~\ref{Fig:ee-Feynman-in} (``set $A$'', initial
state radiation, ISR) and Fig.~\ref{Fig:ee-Feynman-out} (``set $B$'', final
state radiation, FSR), in addition to the well known SM diagrams which are
obtained by substituting the graviton with either a photon or a $Z$ in
diagrams (1) and (2) of sets $A$ and $B$. The SM diagrams are referred to as
``sets $C_\gamma$'', ``$C_Z$'' (both ISR), ``$D_\gamma$'', and ``$D_Z$'' (both
FSR).  It is convenient to separate ISR from FSR since, in the case of ISR,
the graviton propagator does not carry all the momentum of the
electron-positron pair.  In fact, this is the reason the two diagrams labeled
(4) have been classified as ISR and FSR as given in
Figs.~\ref{Fig:ee-Feynman-in} and \ref{Fig:ee-Feynman-out}.
\begin{figure}[htb]
\refstepcounter{figure}
\label{Fig:ee-Feynman-in}
\addtocounter{figure}{-1}
\begin{center}
\setlength{\unitlength}{1cm}
\begin{picture}(12,7.5)
\put(0.5,0.5)
{\mbox{\epsfysize=7.7cm\epsffile{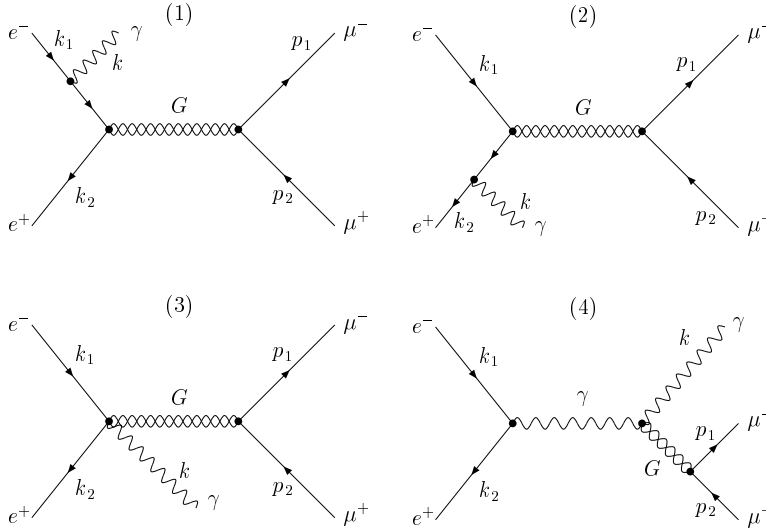}}}
\end{picture}
\vspace*{-8mm}
\caption{Feynman diagrams for ISR in $e^+e^- \to \mu^+\mu^-\gamma$.  We refer
to these diagrams as ``set $A$''.  The corresponding SM diagrams, ``set
$C_\gamma$'' and ``set $C_Z$'', can be obtained by substituting a photon or a
$Z$ for the graviton in diagrams (1) and (2).}
\end{center}
\end{figure}
\begin{figure}[htb]
\refstepcounter{figure}
\label{Fig:ee-Feynman-out}
\addtocounter{figure}{-1}
\begin{center}
\setlength{\unitlength}{1cm}
\begin{picture}(12,7.5)
\put(0.5,0.5)
{\mbox{\epsfysize=7.7cm\epsffile{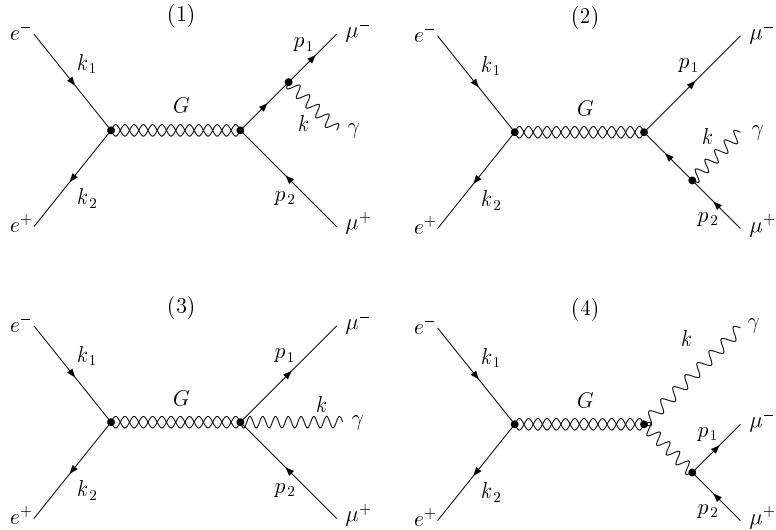}}}
\end{picture}
\vspace*{-8mm}
\caption{Feynman diagrams for FSR. We shall
refer to these as ``set $B$''.  The SM diagrams, ``set $D_\gamma$'' and 
``set $D_Z$'' can be obtained by substituting a photon or a $Z$ for the 
graviton in diagrams (1) and (2).}
\end{center}
\end{figure}

We shall here present the different contributions to the differential cross
section. Let the incident momenta be $k_1$ ($e^-$) and $k_2$ ($e^+$), and the
outgoing momenta be $p_1$ ($\mu^-$), $p_2$ ($\mu^+$) and $k$ ($\gamma$), with
$E_1$, $E_2$ and $\omega$ the corresponding final-state energies.  Then, we
let $x_1$, $x_2$ and $x_3$ denote the fractional energies of the muons and the
photon
\begin{equation}
x_1=E_1/\sqrt{s}, \qquad x_2=E_2/\sqrt{s}, \qquad
x_3=\omega/\sqrt{s}, \qquad 0\le x_i\le\half,
\end{equation}
with $x_1+x_2+x_3=1$. The square of the center of mass energy is $s\equiv
(k_1+k_2)^2=(p_1+p_2+k)^2$ and we denote $s_3\equiv (p_1+p_2)^2=(1-2x_3)s$.
Furthermore, we let $\eta= x_1-x_2$.

As shown in Fig.~\ref{Fig:frame}, we define the scattering angle $\theta$ as
the angle between the incoming electron and the outgoing photon.  When the
polar angle is measured w.r.t.\ the photon momentum (as in
Fig.~\ref{Fig:frame}), the forward--backward asymmetry vanishes. This would
not be the case if we choose a polar angle referring to a muon momentum.
\begin{figure}[htb]
\refstepcounter{figure}
\label{Fig:frame}
\addtocounter{figure}{-1}
\begin{center}
\setlength{\unitlength}{1cm}
\begin{picture}(12,6)
\put(2.5,0.0)
{\mbox{\epsfysize=6cm\epsffile{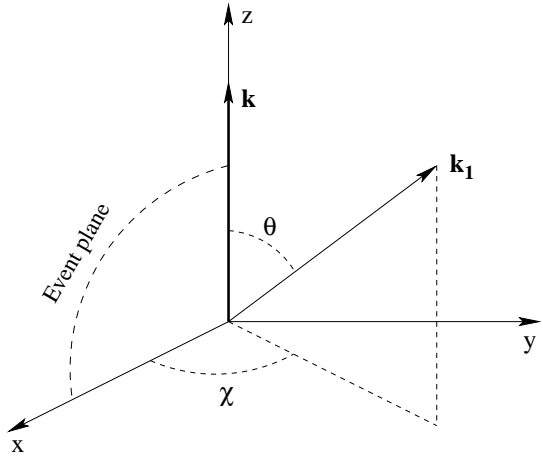}}}
\end{picture}
\caption{Coordinate frame used to describe $e^+e^-\to \mu^+\mu^-\gamma$.  The
incident electron momentum is denoted ${\bf k}_1$, and ${\bf k}$ is the photon
momentum.}
\end{center}
\end{figure}

Following the notation in \cite{Dvergsnes:2002nc}, the different contributions
to the cross section are referred to as
\begin{equation}
\sigma_{ee\to \mu\mu\gamma}=\sigma^{(G)}_{ee\to \mu\mu\gamma}
+\sigma^{(\text{SM})}_{ee\to \mu\mu\gamma}
+\sigma^{(G,\gamma)}_{ee\to \mu\mu\gamma}
+\sigma^{(G,Z)}_{ee\to \mu\mu\gamma},
\end{equation}
where the first term is the graviton contribution (sets $A$ and $B$), the
second term is the Standard-model background (sets $C$ and $D$) and the last
two are graviton--photon and graviton--$Z$ interference terms, respectively.

We shall first consider the graviton exchange diagrams, introducing the
following notation,
\begin{equation}
\sigma^{(G)}_{ee \to \mu\mu\gamma}
=\sigma_{AA}+\sigma_{AB}+\sigma_{BB},
\end{equation}
where $A$ and $B$ refer to the initial- and final-state radiation,
respectively.  The corresponding differential cross section contributions can
now be expressed as
\begin{align}
\label{Eq:dsigma-G}
\frac{d^3\sigma_{AA}}{dx_3 d\eta\,d(\cos\theta)}
&=\frac{\alpha \kappa^4 s \, Q_e^2 }{8192\pi^2}\,
\frac{s_3^2}{(s_3-m_{\vec n}^2)^2+(m_{\vec n}\Gamma_{\vec n})^2}\, 
X_{AA}(x_3,\eta,\cos\theta), \nonumber \\
\frac{d^3\sigma_{AB}}{dx_3 d\eta\,d(\cos\theta)}
&=\frac{\alpha \kappa^4 s\, Q_e Q_\mu}{2048\pi^2}\,
\Re\left[\frac{s_3}{s_3-m_{\vec n}^2-im_{\vec n}\Gamma_{\vec n}}\,
\frac{s}{s-m_{\vec n}^2+im_{\vec n}\Gamma_{\vec n}}\right]\,
X_{AB}(x_3,\eta,\cos\theta), \nonumber \\
\frac{d^3\sigma_{BB}}{dx_3 d\eta\,d(\cos\theta)}
&=\frac{\alpha \kappa^4 s\, Q_\mu^2}{8192\pi^2}\,
\frac{s^2}{(s-m_{\vec n}^2)^2+(m_{\vec n}\Gamma_{\vec n})^2}\,
X_{BB}(x_3,\eta,\cos\theta).
\end{align}
In these expressions, $\alpha$ is the fine-structure constant and
$Q_e=Q_\mu=-1$ is the electron and muon charge. (It is convenient to
distinguish these, in order to more easily trace the origin of the different
terms.) Furthermore, $\kappa$ denotes the strength of the graviton coupling
(to be defined in Sects.~\ref{sec:ADD} and \ref{sec:RS} for the
ADD and RS scenarios), and $m_{\vec n}$ and $\Gamma_{\vec n}$ the mass and
width of the ${\vec n}$'th massive graviton.  The angular distributions, as
well as the way in which the energy is shared by the muons and the photon, are
given by the functions $X_{AA}(x_3,\eta,\cos\theta)$,
$X_{AB}(x_3,\eta,\cos\theta)$ and $X_{BB}(x_3,\eta,\cos\theta)$ defined by
Eq.~(\ref{Eq:XG}) in Appendix~A.  

The denominator of $X_{BB}(x_3,\eta,\cos\theta)$ [see Eq.~(\ref{Eq:XG})]
exhibits the familiar singularities in the infrared and collinear limits, $s_1
\equiv (p_1+k)^2 = s(1-2x_2) \to 0$, $s_2 \equiv (p_2+k)^2 = s(1-2x_1) \to 0$,
as well as a collinear singularity at $s_3 = s(1-2x_3) \to 0$ due to the
fourth Feynman diagram. (Actually, also the ISR contributions in the SM have
this singularity, see Eq.~(\ref{Eq:XSM}), accompanied by a singularity for
small angles.)  The additional singularity means that there is a tendency to
have events with hard photons, like in the analogous hadronic process
\cite{Dvergsnes:2002nc}.

The cross sections for the pure SM background is
\begin{equation}
\sigma^{(\text{SM})}_{ee \to \mu\mu\gamma}
=\sigma_{CC}+\sigma_{CD}+\sigma_{DD},
\end{equation}
where $C$ and $D$ refer to initial- and final-state radiation, with the
corresponding contributions to the differential cross section given by
\begin{align}
\label{Eq:dsigma-SM}
\frac{d^3\sigma_{CC}}{dx_3 d\eta\,d(\cos\theta)}
&=\frac{\alpha^3 Q_e^2}{2s}\, 
{\cal S}_{CC}(s_3,s_3), \nonumber \\
\frac{d^3\sigma_{CD}}{dx_3 d\eta\,d(\cos\theta)}
&=\frac{2\alpha^3 Q_e Q_\mu}{s}\, 
{\cal S}_{CD}(s_3,s), \nonumber \\
\frac{d^3\sigma_{DD}}{dx_3 d\eta\,d(\cos\theta)}
&=\frac{\alpha^3 Q_\mu^2}{2s}\, 
{\cal S}_{DD}(s,s).
\end{align}
Here, the angular and energy distributions are given by
\begin{align}
\label{Eq:sigma-SM}
{\cal S}_{CD}(s_3,s)&=Q_e^2\, Q_\mu^2 
                      X_{C_\gamma D_\gamma}(x_3,\eta,\cos\theta) 
\nonumber \\
&+Q_e\,Q_\mu\, \Re\, \chi(s) X_{C_\gamma D_Z} (x_3,\eta,\cos\theta)
+Q_e\,Q_\mu\, \Re\, \chi(s_3) X_{C_Z D_\gamma}(x_3,\eta,\cos\theta) 
\nonumber \\
&+\Re[\chi^*(s_3)\chi(s)] X_{C_Z D_Z}(x_3,\eta,\cos\theta), 
\end{align}
with ${\cal S}_{CC}(s_3,s_3)$ and ${\cal S}_{DD}(s,s)$ similarly obtained
from Eq.~(\ref{Eq:sigma-SM}) by substituting $(D,s) \leftrightarrow
(C,s_3)$. Furthermore, the $X_{C_\gamma D_\gamma}$ etc.\ are given by
Eq.~(\ref{Eq:XSM}) and the $Z$ propagator is represented by
\begin{equation}
\label{Eq:chi}
\chi(s)=\frac{1}{\sin^2(2\theta_W)}\,
\frac{s}{(s-m_Z^2)+im_Z\Gamma_Z},
\end{equation} 
with $m_Z$ and $\Gamma_Z$ the mass and width of the $Z$ boson, 
and $\theta_W$ the weak mixing angle.  Note that
$\sigma_{C_\gamma C_Z}=\sigma_{C_Z C_\gamma}$ and $\sigma_{D_\gamma
D_Z}=\sigma_{D_Z D_\gamma}$.

For the interference terms between graviton exchange and the SM diagrams,
we introduce the following notation:
\begin{align}
\sigma^{(G,\gamma)}_{ee \to \mu\mu\gamma}
&=\sigma_{AC_\gamma}+\sigma_{BD_\gamma}+\sigma_{AD_\gamma}+\sigma_{BC_\gamma}, 
\nonumber \\[4pt]
\sigma^{(G,Z)}_{ee \to \mu\mu\gamma}
&=\sigma_{AC_Z}+\sigma_{BD_Z}+\sigma_{AD_Z}+\sigma_{BC_Z}.
\end{align}
Like above, the subscripts indicate the diagram sets involved.  The
corresponding differential cross section contributions are given by
\begin{align} \label{Eq:dsigma-G-SM}
\frac{d^3\sigma_{AC_\gamma}}{dx_3 d\eta\,d(\cos\theta)}
&=\frac{\alpha^2 \kappa^2\, Q_e^3 Q_\mu}{32\pi}\,
\Re\left[\frac{s_3}{s_3-m_{\vec n}^2+im_{\vec n}\Gamma_{\vec n}}\right]\, 
X_{AC_\gamma}(x_3,\eta,\cos\theta), \nonumber \\
\frac{d^3\sigma_{AC_Z}}{dx_3 d\eta\,d(\cos\theta)}
&=\frac{\alpha^2 \kappa^2\, Q_e^2}{64\pi}\,
\Re\left[\chi^*(s_3)\frac{s_3}{s_3-m_{\vec n}^2+im_{\vec n}\Gamma_{\vec n}}
\right]\, 
X_{AC_Z}(x_3,\eta,\cos\theta), \nonumber \\
\frac{d^3\sigma_{BD_\gamma}}{dx_3 d\eta\,d(\cos\theta)}
&=\frac{\alpha^2 \kappa^2\, Q_e Q_\mu^3}{32\pi}\,
\Re\left[\frac{s}{s-m_{\vec n}^2+im_{\vec n}\Gamma_{\vec n}}\right]\, 
X_{BD_\gamma}(x_3,\eta,\cos\theta), \nonumber \\
\frac{d^3\sigma_{BD_Z}}{dx_3 d\eta\,d(\cos\theta)}
&=\frac{\alpha^2 \kappa^2\, Q_\mu^2}{64\pi}\,
\Re\left[\chi^*(s)\frac{s}{s-m_{\vec n}^2+im_{\vec n}\Gamma_{\vec n}}
\right]\, X_{BD_Z}(x_3,\eta,\cos\theta), \nonumber \\
\frac{d^3\sigma_{AD_\gamma}}{dx_3 d\eta\,d(\cos\theta)}
&=\frac{\alpha^2 \kappa^2\, Q_e^2 Q_\mu^2}{128\pi}\,
\Re\left[\frac{s_3}{s_3-m_{\vec n}^2+im_{\vec n}\Gamma_{\vec n}}\right]\, 
X_{AD_\gamma}(x_3,\eta,\cos\theta), \nonumber \\
\frac{d^3\sigma_{AD_Z}}{dx_3 d\eta\,d(\cos\theta)}
&=\frac{\alpha^2 \kappa^2\, Q_e Q_\mu}{128\pi}\,
\Re\left[\chi^*(s)\frac{s_3}{s_3-m_{\vec n}^2+im_{\vec n}\Gamma_{\vec n}}
\right]\, X_{AD_Z}(x_3,\eta,\cos\theta), \nonumber \\
\frac{d^3\sigma_{BC_\gamma}}{dx_3 d\eta\,d(\cos\theta)}
&=\frac{\alpha^2 \kappa^2\, Q_e^2 Q_\mu^2}{128\pi}\,
\Re\left[\frac{s}{s-m_{\vec n}^2+im_{\vec n}\Gamma_{\vec n}}\right]\, 
X_{BC_\gamma}(x_3,\eta,\cos\theta), \nonumber \\
\frac{d^3\sigma_{BC_Z}}{dx_3 d\eta\,d(\cos\theta)}
&=\frac{\alpha^2 \kappa^2\, Q_e Q_\mu}{128\pi}\,
\Re\left[\chi^*(s_3)\frac{s}{s-m_{\vec n}^2+im_{\vec n}\Gamma_{\vec n}}
\right]\, 
X_{BC_Z}(x_3,\eta,\cos\theta).
\end{align}
The $X_{AC_\gamma}$ etc.\ are given in Appendix~A.

An overview of the notations used for the different contributions to the cross
section is given in Table~\ref{tab:ee-table}.
\begin{table}[htb]
\refstepcounter{table}
\addtocounter{table}{-1}
\label{tab:ee-table}
\begin{center}
\setlength{\unitlength}{1cm}
\begin{picture}(8,8)
\put(0.,-0.3)
{\mbox{\epsfysize=9.0cm\epsffile{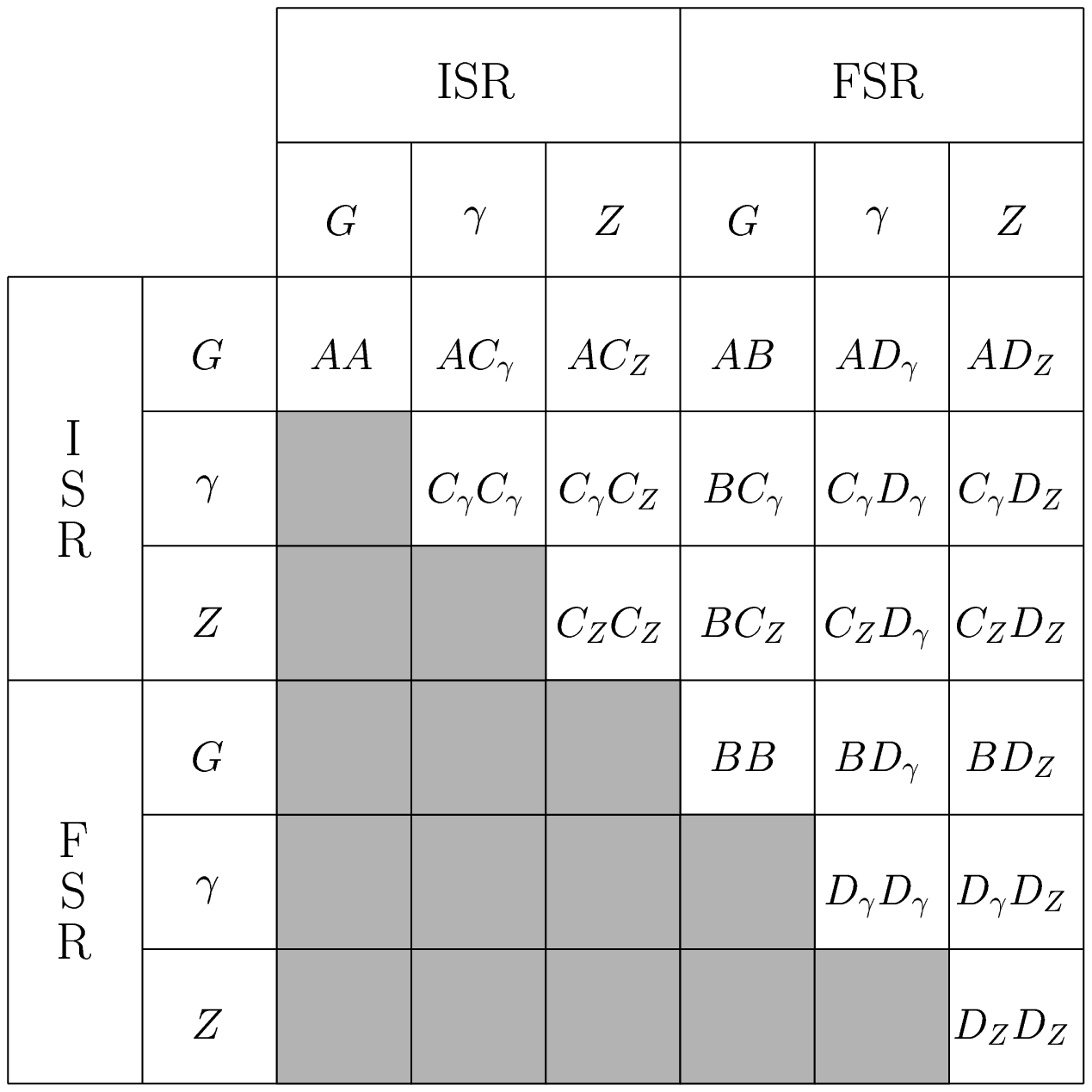}}}
\end{picture}
\end{center}
\vspace*{-4pt}
\caption{Notation used for different combinations of amplitudes.  Compare the
labeling of diagrams in Figs.~\ref{Fig:ee-Feynman-in} and
\ref{Fig:ee-Feynman-out}.}
\end{table}

\subsection{Total cross section}
To obtain the total cross section, we integrate the differential cross
section presented in Sec.~\ref{sec:diff-sigma} within the following limits:
\begin{equation}
\label{Eq:sigma-tot}
\sigma_{ee \to \mu\mu\gamma}=
\int_{-1+c_\text{cut}}^{1-c_\text{cut}} d(\cos\theta) 
\int_{x_3^\text{min}}^{x_3^\text{max}} dx_3
\int_{-x_3 + y_\text{cut}}^{x_3 - y_\text{cut}} d\eta
\frac{d^3\sigma_{ee \to \mu\mu\gamma}}{dx_3 d\eta\,d(\cos\theta)}.
\end{equation}
Since the detector has a `blind' region very close to the beam pipe, we impose
a cut, $|\cos\theta|<1-c_\text{cut}$, with $c_\text{cut}=0.005$, which
translates into a lower bound on $\sin\theta_\text{min}\simeq 0.1$ or an
angular cut of $\theta_\text{min} \simeq 100$~mrad. This cut removes the
singularity due to initial-state radiation (recall that $\theta$ is the angle
between the photon and the incident beam).

The resolution cut, $y_\text{cut}=0.005$, is imposed to exclude collinear
events, i.e., by requiring $s_i=(1-2x_i)s>y_\text{cut}\,s$. For fixed $x_3$,
this leads to $|\eta|<x_3-y_\text{cut}$, where $\eta=x_1-x_2$.  The variable
$x_3$ is bounded by the allowed values of $s_i$, giving
$y_\text{cut}<x_3<\half(1-y_\text{cut}) \equiv x_3^\text{max}$.

As a result of the cut on $s_i$, the minimal photon momentum is
$k_\text{min}=y_\text{cut}\sqrt{s}$. For $\sqrt{s}=500$~GeV and the chosen
value for $y_\text{cut}$, this becomes 2.5~GeV. In addition to this cut we
shall also require that the photon perpendicular momentum is subject to an
absolute cut, $k_\perp=k\sin\theta>k_\perp^\text{min}$.  Here we choose
$k_\perp^\text{min}=\xi_\text{cut}\sqrt{s}$, with $\xi_\text{cut}=0.005$. For
$\sqrt{s}=500$~GeV, $k_\perp^\text{min}=2.5$~GeV, which means that photons
with momentum $k_\text{min}$ only survive this cut when $\sin\theta=1$. If
$\sin\theta=\sin\theta_\text{min}$, only photons of $k>25$~GeV survive the
cuts.

When expressed in terms of the variables $x_3$ and $\cos\theta$, the $k_\perp$
constraint becomes $x_3\sqrt{1-\cos^2\theta}>\xi_\text{cut}$. Thus, for a
given $\cos\theta$ in the allowed range, we find
\begin{equation}
x_3>x_3^\text{min}=\text{max}
\left(\frac{\xi_\text{cut}}{\sqrt{1-\cos^2\theta}},y_\text{cut}\right)
\end{equation}

In order to exclude radiative return to the $Z$, we will also consider the cut
\begin{equation} \label{Eq:rr-cut}
s_3>(m_Z+3\Gamma_Z)^2 \equiv y_\text{cut}^\text{rr} s. 
\end{equation}
This implies
\begin{equation}
y_\text{cut}^\text{rr}=\frac{m_Z^2}{s}\left(1+\frac{3\Gamma_Z}{m_Z}\right)^2
\simeq 1.17 \times\frac{m_Z^2}{s},
\end{equation}
which for $\sqrt{s}=500$~GeV gives $y_\text{cut}^\text{rr} \simeq 0.039$.
This value will modify the upper bound $x_3^\text{max}$, which will become
$\half(1-y_\text{cut}^\text{rr})$, but not affect the lower bound,
$x_3^\text{min}$, nor the limits on $\eta$.
\subsection{Photon perpendicular momentum distribution}
It is instructive to consider the spectrum of the photon perpendicular
momentum, $k_\perp$, since this has no analogue in the two-body final state
process. As anticipated above, we expect it to be harder than in the QED
case. The relevant differential cross sections can be obtained from the
expressions in Sec.~\ref{sec:diff-sigma} upon a change of variables from
$(x_3, \cos\theta) \to (k_\perp, k_\|)$. From the definitions,
$k_\perp=\sqrt{s}x_3\sin\theta$ and $k_\|=\sqrt{s}x_3\cos\theta$, we get
$dx_3d(\cos\theta)\to|J|dk_\|dk_\perp$ with the Jacobian
\begin{equation}
|J|=\frac{k_\perp}{\sqrt{s}k^2}=\frac{k_\perp}{\sqrt{s}(k_\perp^2+k_\|^2)}.
\end{equation}

The photon perpendicular momentum spectrum is now obtained from
\begin{equation}
\label{Eq:sigma-kperp}
\frac{d\sigma_{ee \to \mu\mu\gamma}}{dk_\perp}
=\int_{-k_\|^\text{max}}^{k_\|^\text{max}} dk_\| 
\int_{-x_3 + y_\text{cut}}^{x_3 - y_\text{cut}} d\eta\,
\frac{d\sigma^3_{ee \to \mu\mu\gamma}}{dk_\perp dk_\| d\eta}.
\end{equation}
Given some $k_\perp$ within the allowed region
$\xi_\text{cut}\sqrt{s}<k_\perp<\frac{\sqrt{s}}{2}(1-y_\text{cut})$, we find
\begin{equation}
|k_\||<k_\|^\text{max}=\text{min}\left(\sqrt{\frac{s}{4}(1-y_\text{cut})^2 
- k_\perp^2},\frac{\sqrt{s}}{2}(1-y_\text{cut})(1-c_\text{cut})\right).
\end{equation}
The resolution cut, $y_\text{cut}$, and also the radiative-return cut,
$y_\text{cut}^\text{rr}$, will be the same as for the total cross section, and
the radiative-return cut will affect both $k_\perp^\text{max}$ and
$k_\|^\text{max}$.
\subsection{Photon angular distribution}
For the two-body final states $e^+e^-\to\mu^+\mu^-$ and
$e^+e^-\to\gamma\gamma$, the QED angular distributions are given by the
familiar $1+\cos^2\theta$ and $(1+\cos^2\theta)/(1-\cos^2\theta)$. For
graviton exchange, the corresponding distributions become
$1-3\cos^2\theta+4\cos^4\theta$ and $1-\cos^4\theta$ (see e.g.\
\cite{Cheung:1999wt}). In both these cases, the higher powers are due to the
spin-2 coupling.  For the three-body case, we get similar expressions (see the
Appendix).  Note that the ISR contribution has a structure similar to that of
the diphoton channel, with a $1-\cos^2\theta$ singularity in the denominator,
whereas graviton exchange gives quartic terms in $\cos\theta$.  

In order to emphasize the photons originating from graviton exchange over
those from the collinear singularities (dominated by the SM contributions), we
will here consider the angular distribution of the {\it photon} with respect
to the incident beam:
\begin{equation}
\label{Eq:sigma-ang}
\frac{d\sigma_{ee \to \mu\mu\gamma}}{d(\cos\theta)}
=\int_{x_3^\text{min}}^{x_3^\text{max}} dx_3 
\int_{-x_3 + y_\text{cut}}^{x_3 - y_\text{cut}} d\eta
\frac{d^3\sigma_{ee \to \mu\mu\gamma}}{dx_3 d\eta\,d(\cos\theta)},
\end{equation}
with the cuts as given above.

\section{The ADD scenario} \label{sec:ADD}
We first turn our attention to the ADD scenario \cite{Arkani-Hamed:1998rs},
where there is essentially a continuum of massive graviton states up to some
cut-off $M_S$, where a more fundamental theory, presumably low-scale string
physics, takes over.  Following the convention of \cite{Han:1999sg}, the
coherent sum over all KK modes in a tower is performed by substituting for the
sum over graviton propagators the following expression:
\begin{equation}
\label{Eq:propagatorsum}
\frac{\kappa^2}{s-m_{\vec n}^2+im_{\vec n} \Gamma_{\vec n}} 
\equiv -i\kappa^2 D(s)
\underset{\sum_{\vec n}}{\longrightarrow}
\frac{8\pi s^{n/2-1}}{M_S^{n+2}}\, [2I(M_S/\sqrt{s}) -i\pi],
\end{equation}
with
\begin{equation}
\label{Eq:III}
I(M_S/\sqrt{s})=
\begin{cases}
\begin{displaystyle}
-\sum_{k=1}^{n/2-1}\frac{1}{2k}\, \left(\frac{M_S}{\sqrt{s}}\right)^{2k} 
- \frac{1}{2}\log\left(\frac{M_S^2}{s} - 1\right), 
\end{displaystyle}
& n=\text{even}, \\
\begin{displaystyle}
- \sum_{k=1}^{(n-1)/2}\frac{1}{2k-1} 
\left(\frac{M_S}{\sqrt{s}}\right)^{2k-1} 
+ \frac{1}{2}\log\left(\frac{M_S + \sqrt{s}}{M_S - \sqrt{s}}\right), 
\end{displaystyle}
& n=\text{odd}.
\end{cases}
\end{equation}
for $n$ extra dimensions. 

Since the role of higher-order loop effects is rather unknown
\cite{Contino:2001nj}, these expressions should not be taken too literally.
However, in order to preserve the qualitative difference between the two
propagators $D(s)$ and $D(s_3)$ (see Eq.~(\ref{Eq:propagatorsum})), and thus
more easily keep track of the contributions of different Feynman diagrams, we
shall use the expressions of Eq.~(\ref{Eq:III}).  In the approach of
\cite{Giudice:1999ck} and \cite{Hewett:1999sn} the $n$-dependence is absorbed
in the cut-off so that $D(s)$ and $D(s_3)$ are indistinguishable. For $n=4$
and $M_S\gg\sqrt{s}$, the cut-off $M_S$ is comparable to $\Lambda_T$ of
ref.~\cite{Giudice:1999ck} and $M_H$ of \cite{Hewett:1999sn}.

\subsection{Total cross sections}

\begin{figure}[htb]
\refstepcounter{figure}
\label{Fig:add-sigtot-tesla}
\addtocounter{figure}{-1}
\begin{center}
\setlength{\unitlength}{1cm}
\begin{picture}(14,6.5)
\put(0,0.0)
{\mbox{\epsfysize=7cm\epsffile{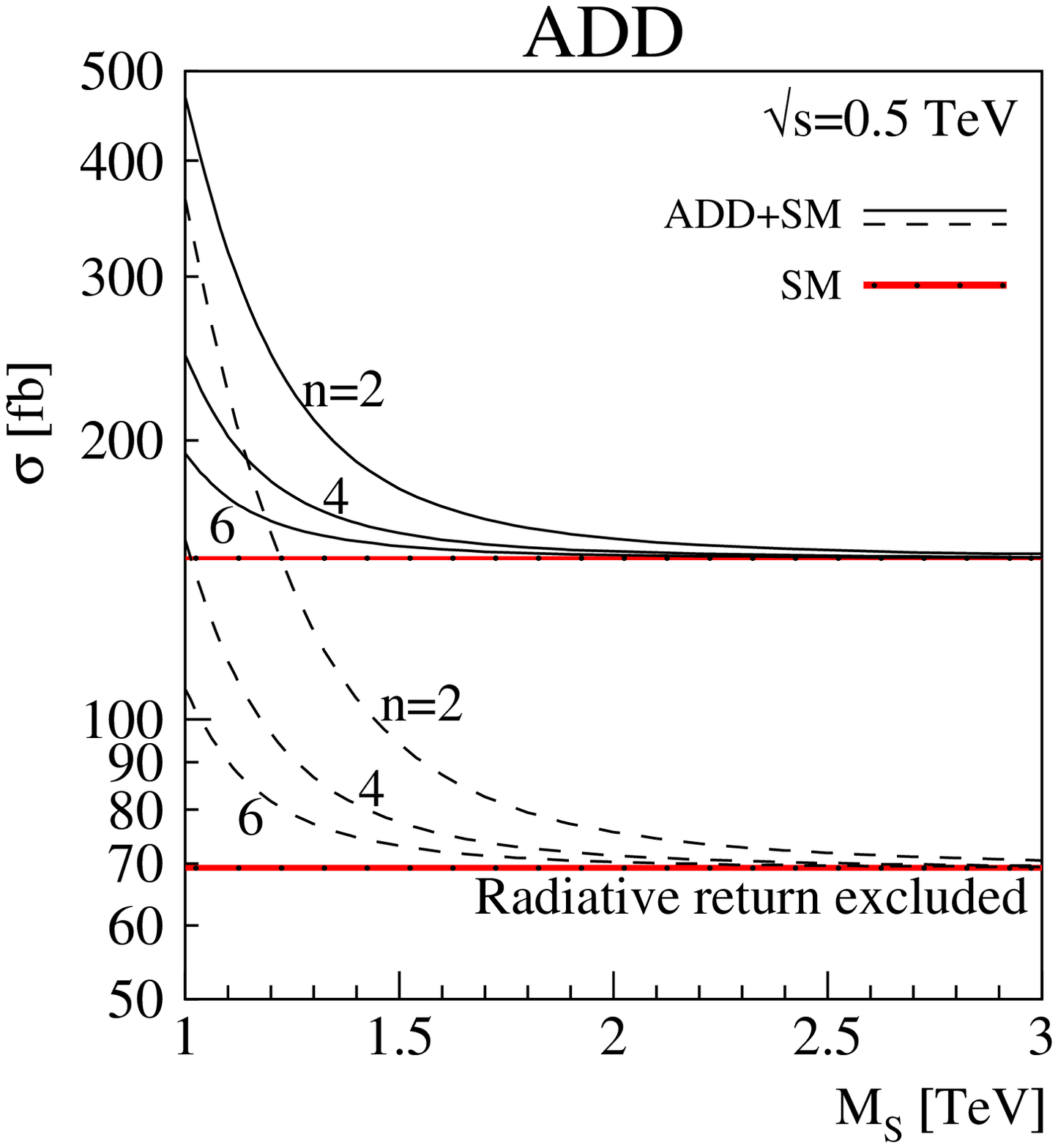}}
 \mbox{\epsfysize=7cm\epsffile{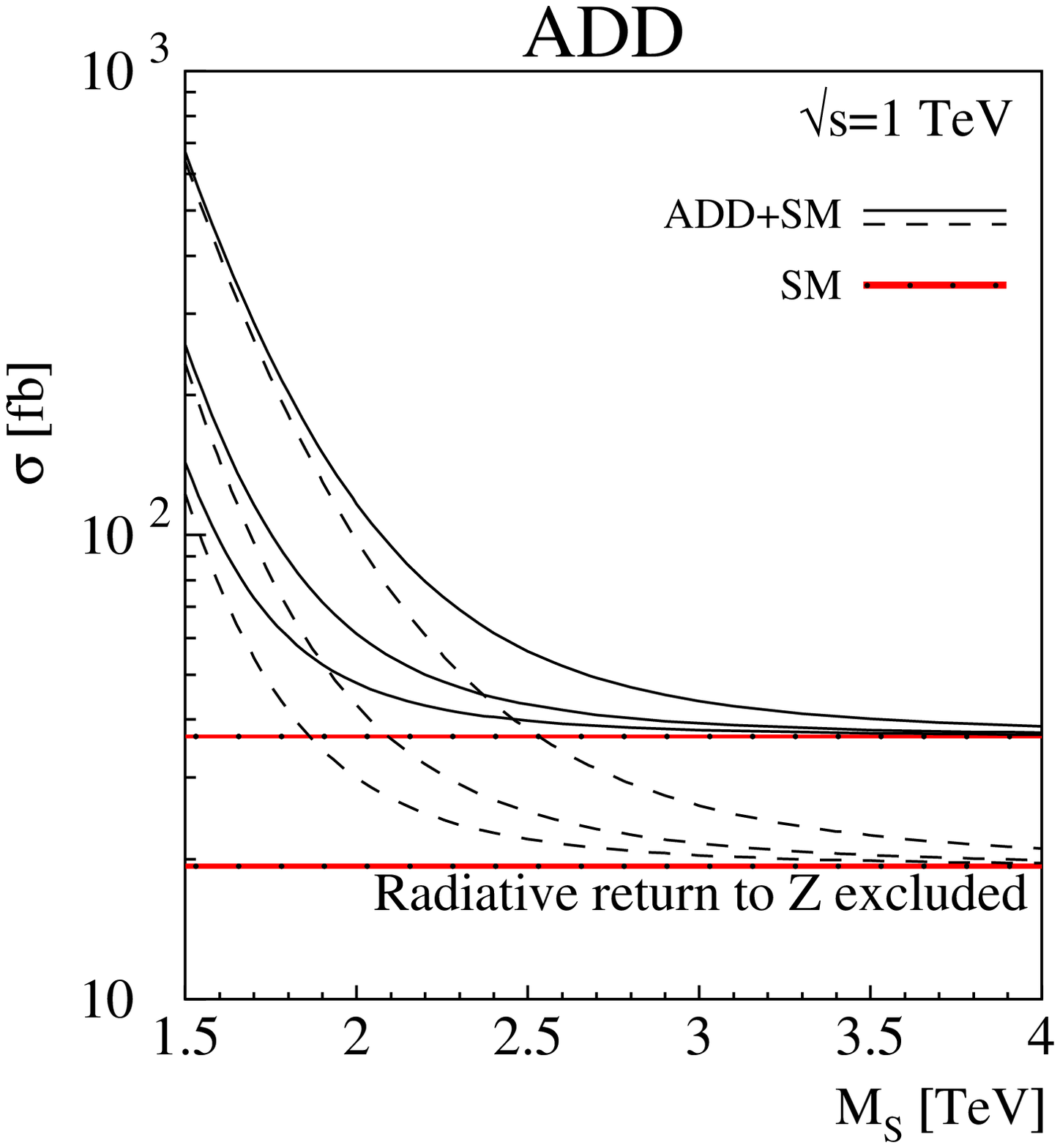}}}
\end{picture}
\vspace*{-8mm}
\caption{Total cross sections for $e^+e^-\to\mu^+\mu^-\gamma$ vs.\ $M_S$, for
$\sqrt{s}=0.5$ and 1~TeV, and $n=2$, 4 and 6, with (solid) and without
(dashed) radiative return to the $Z$ pole.  The SM value is represented by a
band corresponding to $\Lumint=300~\text{fb}^{-1}$.}
\end{center}
\end{figure}

\begin{figure}[htb]
\refstepcounter{figure}
\label{Fig:add-sigtot-clic}
\addtocounter{figure}{-1}
\begin{center}
\setlength{\unitlength}{1cm}
\begin{picture}(14,6.7)
\put(0,0.3)
{\mbox{\epsfysize=7cm\epsffile{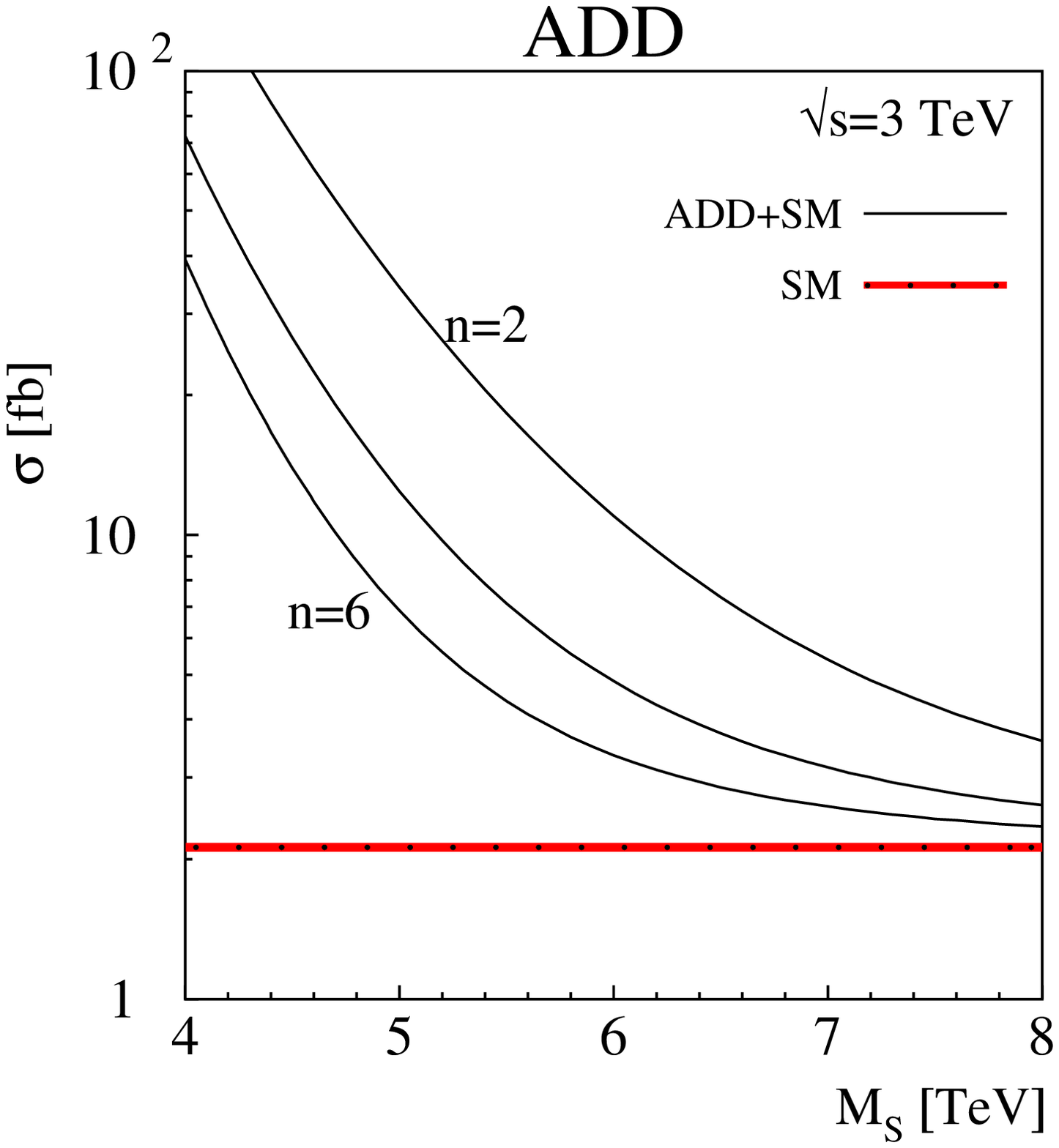}}
 \mbox{\epsfysize=7cm\epsffile{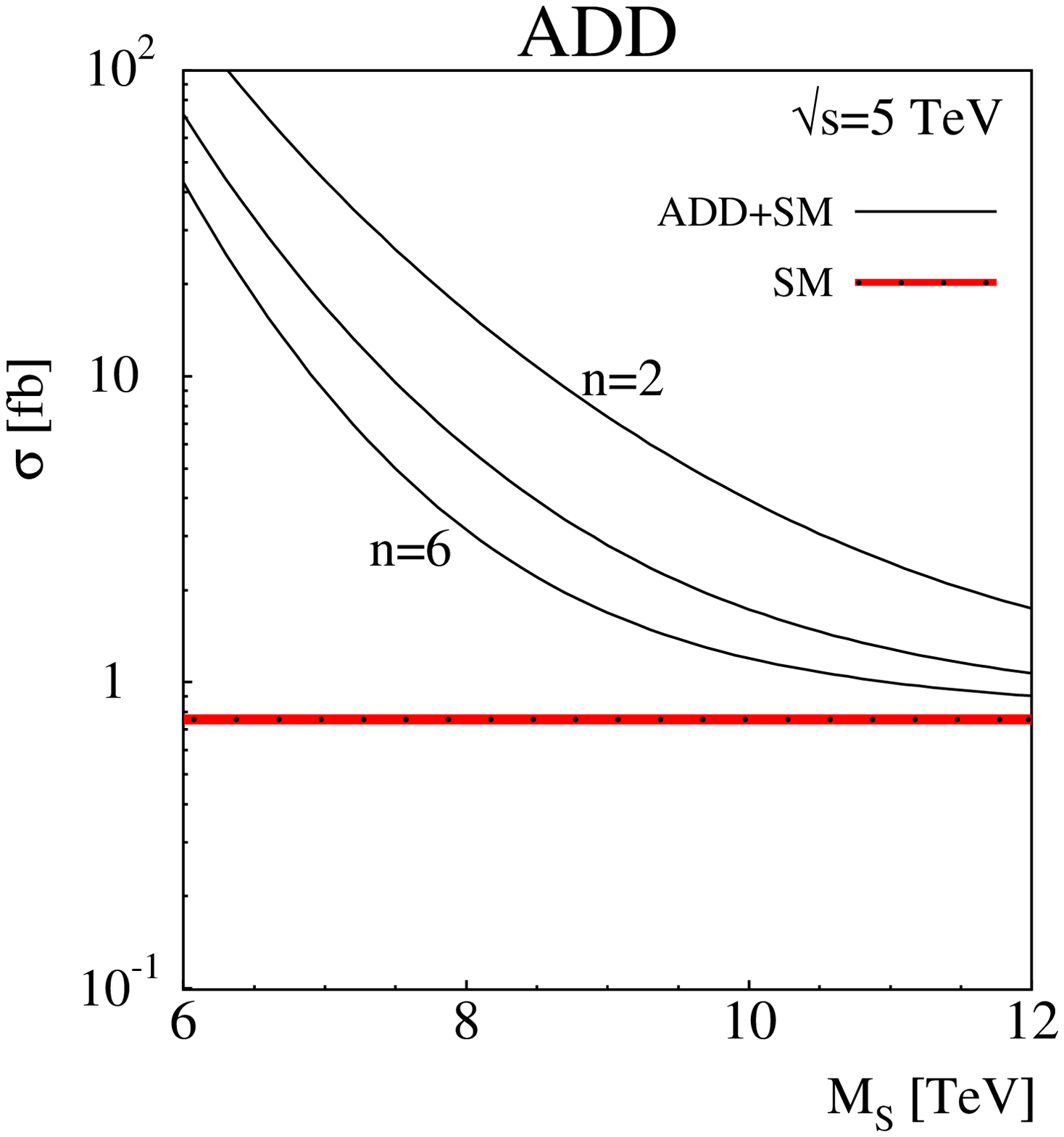}}}
\end{picture}
\vspace*{-8mm}
\caption{Total cross sections for $e^+e^-\to\mu^+\mu^-\gamma$ vs.\ $M_S$, for
$\sqrt{s}=3$ and 5~TeV, and $n=2$, 4, and 6.  The SM value is represented by a
band corresponding to $\Lumint=1000~\text{fb}^{-1}$.}
\end{center}
\end{figure}

In Figs.~\ref{Fig:add-sigtot-tesla} and \ref{Fig:add-sigtot-clic}, we present
the total cross section [see Eq.~(\ref{Eq:sigma-tot})] vs.\ the UV cut-off
$M_S$, for $n=2$, 4 and 6.  (For $n=2$, this range of $M_S$ is actually in
conflict with astrophysical data \cite{Hannestad:2001jv}.) Different collider
energies are considered, $\sqrt{s}=0.5$ and $1.0$~TeV in
Fig.~\ref{Fig:add-sigtot-tesla}, and $3.0$ and $5.0$~TeV in
Fig.~\ref{Fig:add-sigtot-clic}. For $\sqrt{s}=3$ and $5~\text{TeV}$, radiative
return to $Z$ is already excluded by the $y$ cut, and therefore only one set
of curves is shown.

It is seen that the integrated cross sections can have a significant
enhancement over the SM result provided $M_S$ is not too much above the actual
c.m.\ energy.  Also, we note that removing the radiative return to the $Z$
according to the criterion (\ref{Eq:rr-cut}), the cross section is reduced
significantly.  Since this mostly affects the SM background, the relative
magnitude of the ``signal'' increases.

As a rough indication of the precision to be expected, we display the
$1\sigma$ statistical error band around the SM values, corresponding to an
integrated luminosity of $300~\text{fb}^{-1}$ for the cases of $\sqrt{s}=0.5$
and 1~TeV, and $1000~\text{fb}^{-1}$ for $\sqrt{s}=3$ and 5~TeV (we take the
efficiency to be 1 throughout the paper).  We note that the sensitivity of the
integrated cross section extends out to values of $\sqrt{s}$ that are a few
times the available c.m.\ energy.  However, since it is a higher-order
process, suppressed by a factor of the order $\alpha/\pi$, the sensitivity
does not compete with that of the two-body final states
\cite{Hewett:1999sn,Rizzo:2002pc,Osland:2003fn}.
\subsection{Photon perpendicular momentum distributions}

Because of the Feynman diagrams (3) and (4), the photon tends to be harder
than in QED or the SM \cite{Dvergsnes:2002nc}. This is illustrated in
Fig.~\ref{Fig:add-kperp} for $\sqrt{s}=0.5~\text{TeV}$, where we show
$d\sigma_{ee \to \mu\mu\gamma}/dk_\perp$ as given by
Eq.~(\ref{Eq:sigma-kperp}) for $n=4$ and $M_S=1.5~\text{TeV}$.  The peak at
the highest values of $k_\perp\sim\half\sqrt{s}$ is due to radiative return to
the $Z$. As can be seen in this figure, radiative return mainly affects the SM
background, and can be removed by a cut on $s_3$ [see Eq.~(\ref{Eq:rr-cut})].

\begin{figure}[htb]
\refstepcounter{figure}
\label{Fig:add-kperp}
\addtocounter{figure}{-1}
\begin{center}
\setlength{\unitlength}{1cm}
\begin{picture}(14,6.8)
\put(2,0.3)
{\mbox{\epsfysize=7.2cm\epsffile{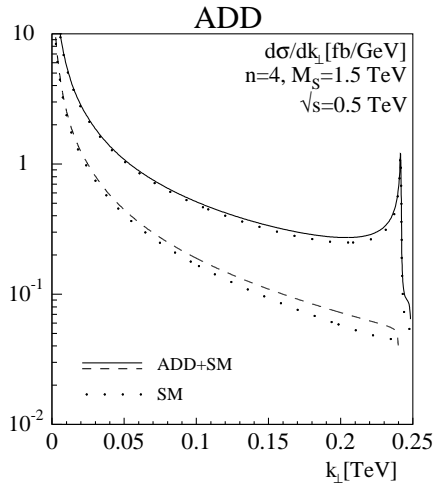}}}
\end{picture}
\vspace*{-8mm}
\caption{Photon perpendicular momentum distribution for $n=4$, with (upper)
and without (lower curve) radiative return to $Z$. The SM contribution is
dotted.}
\end{center}
\end{figure}

\begin{figure}[htb]
\refstepcounter{figure}
\label{Fig:add-kperp-tesla-bin}
\addtocounter{figure}{-1}
\begin{center}
\setlength{\unitlength}{1cm}
\begin{picture}(14,6.5)
\put(-1,0.0)
{\mbox{\epsfysize=7.2cm\epsffile{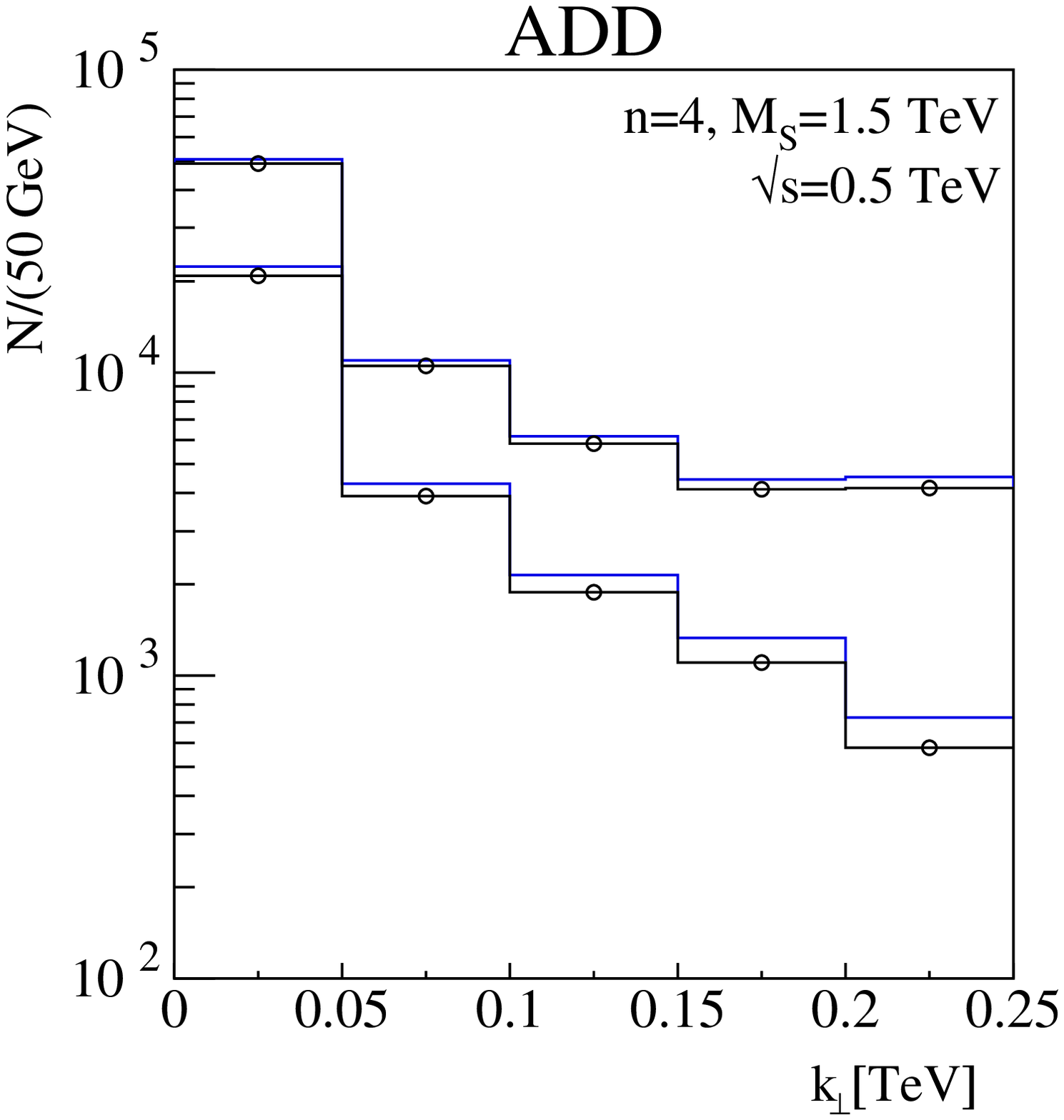}}
 \mbox{\epsfysize=7.2cm\epsffile{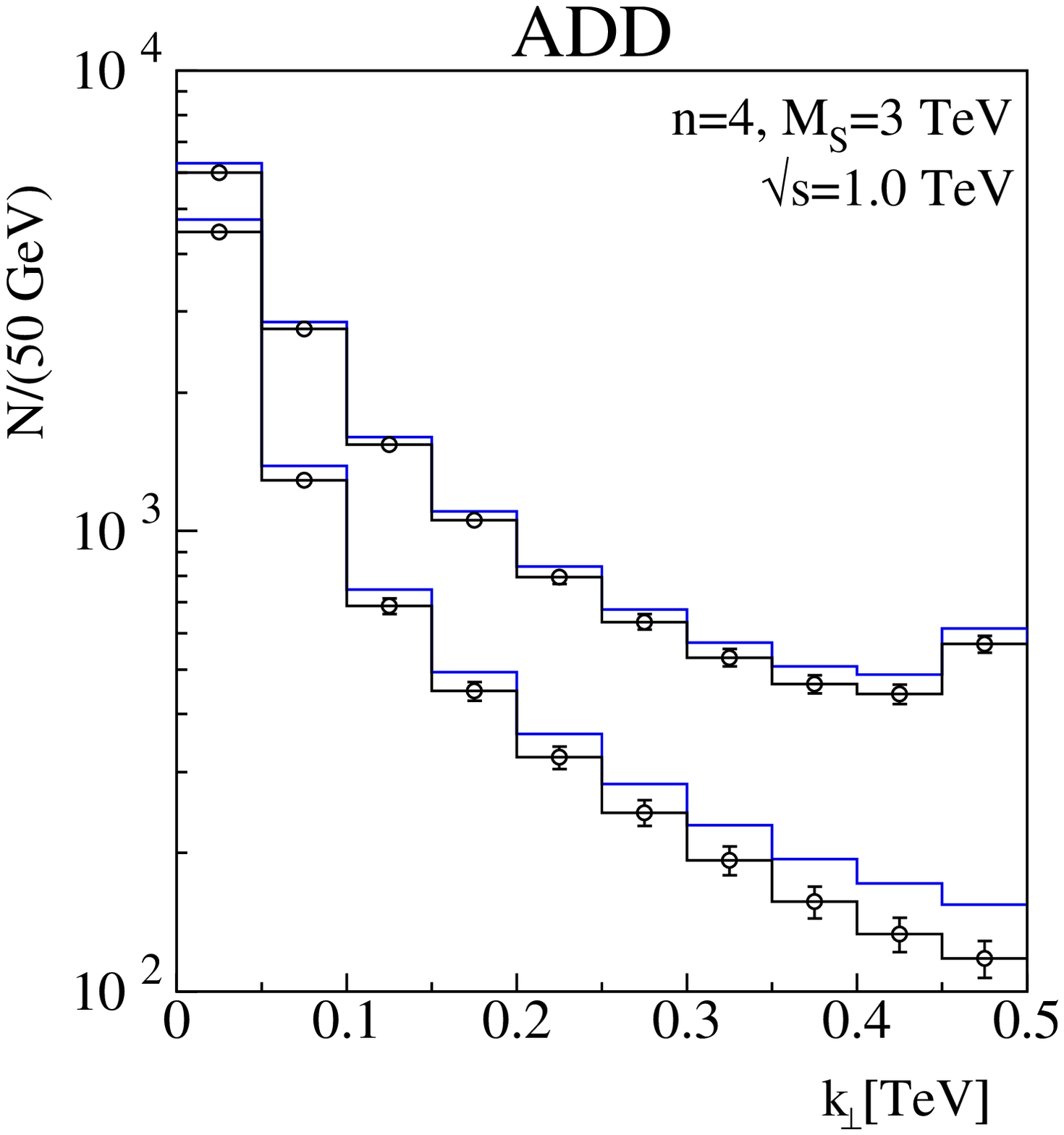}}}
\end{picture}
\vspace*{-8mm}
\caption{Photon perpendicular momentum distributions for $n=4$, with (upper)
and without (lower set of curves) radiative return to $Z$. The SM contribution
is displayed with error bars (invisible in the left panel) corresponding to
$300~\text{fb}^{-1}$.}
\end{center}
\end{figure}

\begin{figure}[htb]
\refstepcounter{figure}
\label{Fig:add-kperp-clic-bin}
\addtocounter{figure}{-1}
\begin{center}
\setlength{\unitlength}{1cm}
\begin{picture}(14,6.5)
\put(-1,0.0)
{\mbox{\epsfysize=7.2cm\epsffile{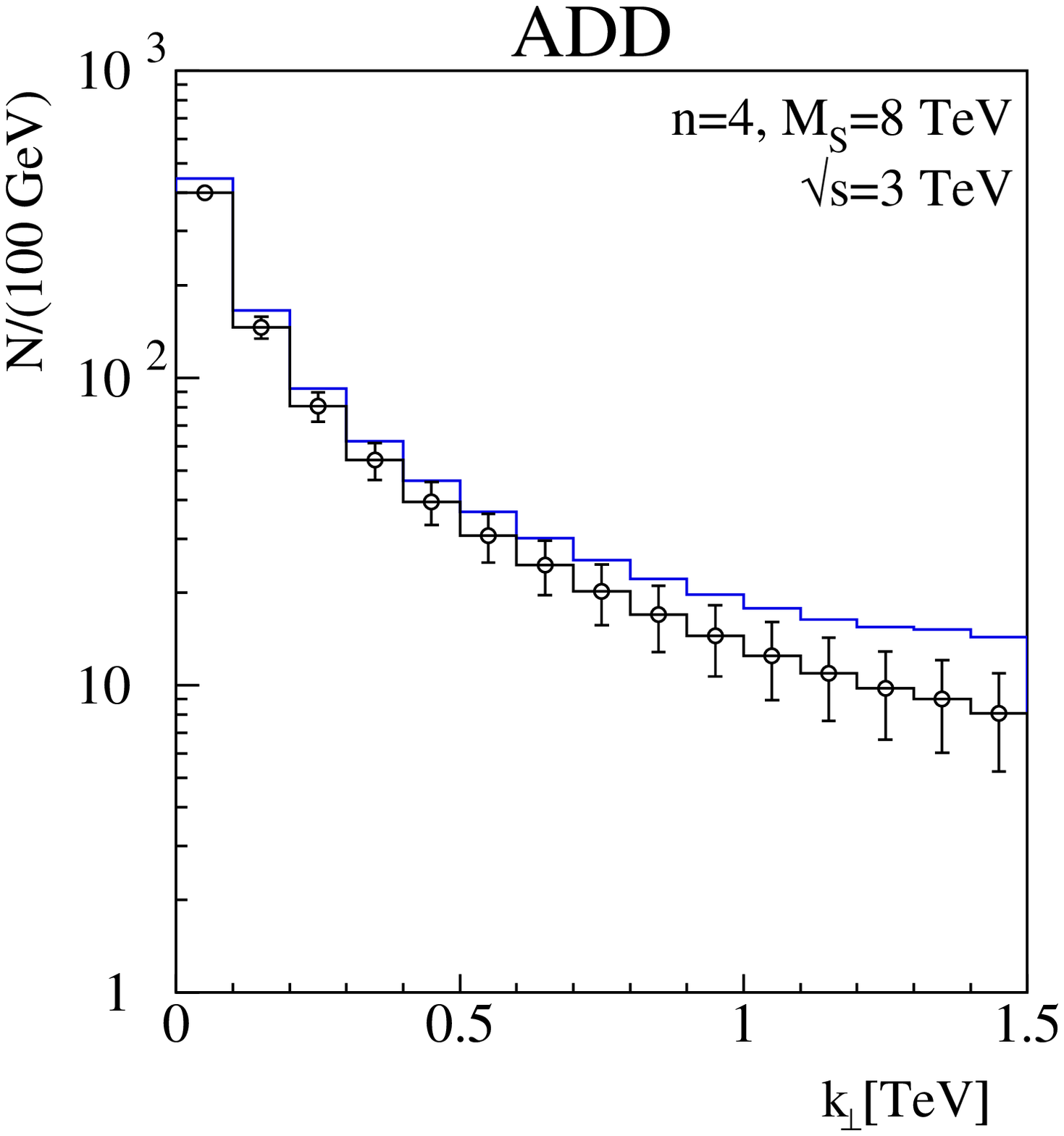}}
 \mbox{\epsfysize=7.2cm\epsffile{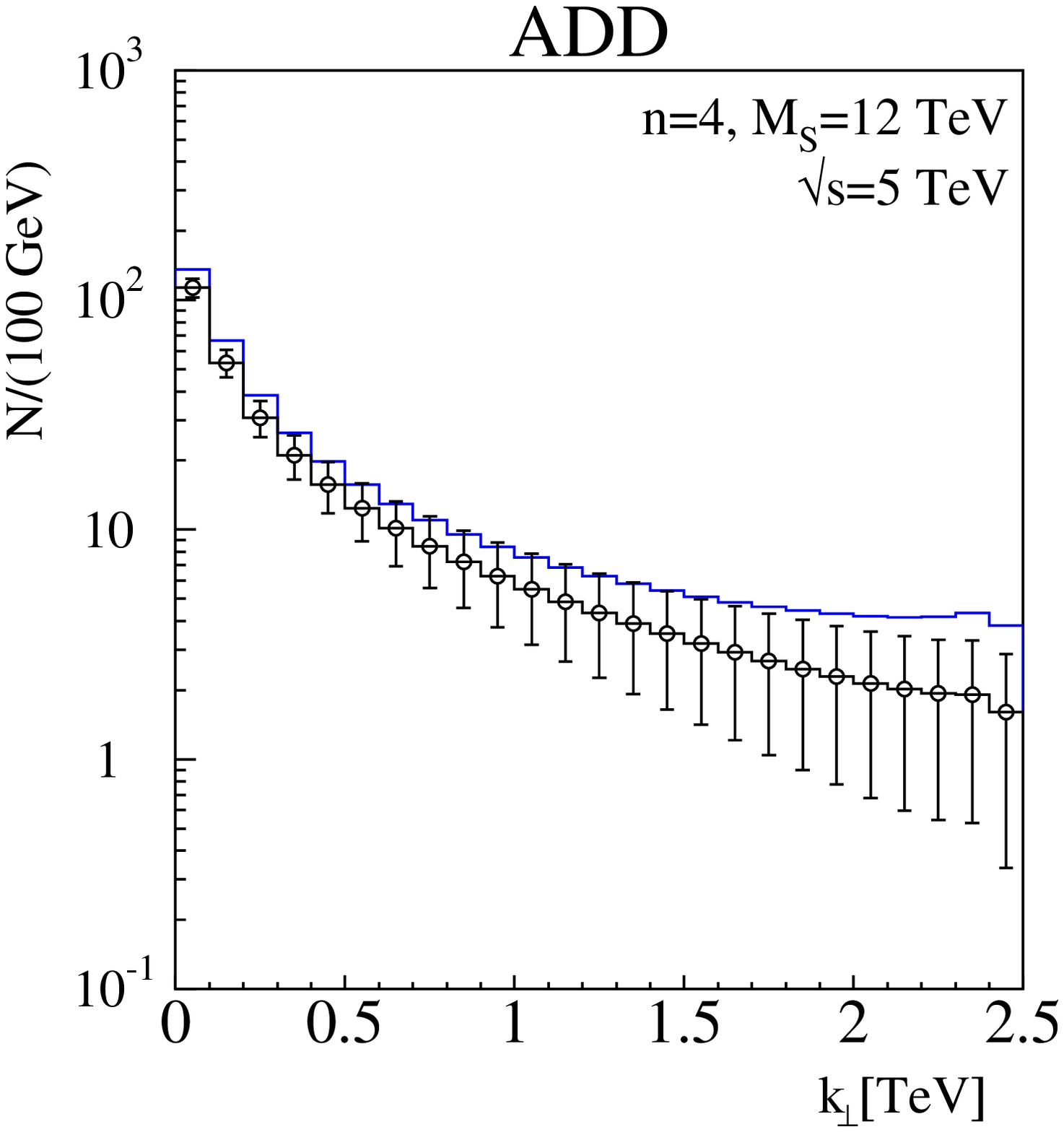}}}
\end{picture}
\vspace*{-8mm}
\caption{Photon perpendicular momentum distributions for $n=4$. The SM
contribution is displayed with error bars corresponding to
$1000~\text{fb}^{-1}$.}
\end{center}
\end{figure}

In order to give an idea how significant the difference is, we also show in
Figs.~\ref{Fig:add-kperp-tesla-bin} and \ref{Fig:add-kperp-clic-bin}
bin-integrated $k_\perp$ distributions, corresponding to an integrated
luminosity of $300~\text{fb}^{-1}$ for 0.5 and 1~TeV, with a bin width of
50~GeV, and an integrated luminosity of $1000~\text{fb}^{-1}$ for 3 and 5~TeV,
with a bin width of 100~GeV. In these figures, we have taken $n=4$ and
selected values of $M_S$, namely 1.5, 3, 8 and 12~TeV.

It is seen that, after the binning in $k_\perp$, the excess of the ADD+SM
cross section over the SM cross section remains significant for the considered
luminosities. As anticipated, the excess increases with $k_\perp$, also with
respect to the statistical uncertainty, in particular after the removal of
radiative-return events. The quantitative benefit of this radiative-return cut
will of course depend on the integrated luminosity and the cut parameter [see
Eq.~(\ref{Eq:rr-cut})] as well as on $M_S$.  As mentioned above, for
$\sqrt{s}=3$ and $5~\text{TeV}$, radiative return to $Z$ is already excluded
by the $y$ cut, thus only one set of curves is displayed.

\subsection{Photon angular distributions}
Due to conventional ISR (diagrams (1) and (2) in
Fig.~\ref{Fig:ee-Feynman-in}), the photon angular distributions are peaked
near the beam direction. This is the case for any $s$-channel exchange, and
stems from the collinear singularity of those diagrams.  Similarly, diagrams
(1) and (2) in Fig.~\ref{Fig:ee-Feynman-out} (final-state radiation) lead
predominantly to photons close to the directions of the final-state muon
momenta.  On the other hand, the diagrams (3) and (4), for ISR as well as for
FSR, do not have such collinear singularities, and could therefore lead to
distinctive features, different from those of the SM.
\begin{figure}[htb]
\refstepcounter{figure}
\label{Fig:add-angular}
\addtocounter{figure}{-1}
\begin{center}
\setlength{\unitlength}{1cm}
\begin{picture}(14,6.5)
\put(-1,0.0)
{\mbox{\epsfysize=7.2cm\epsffile{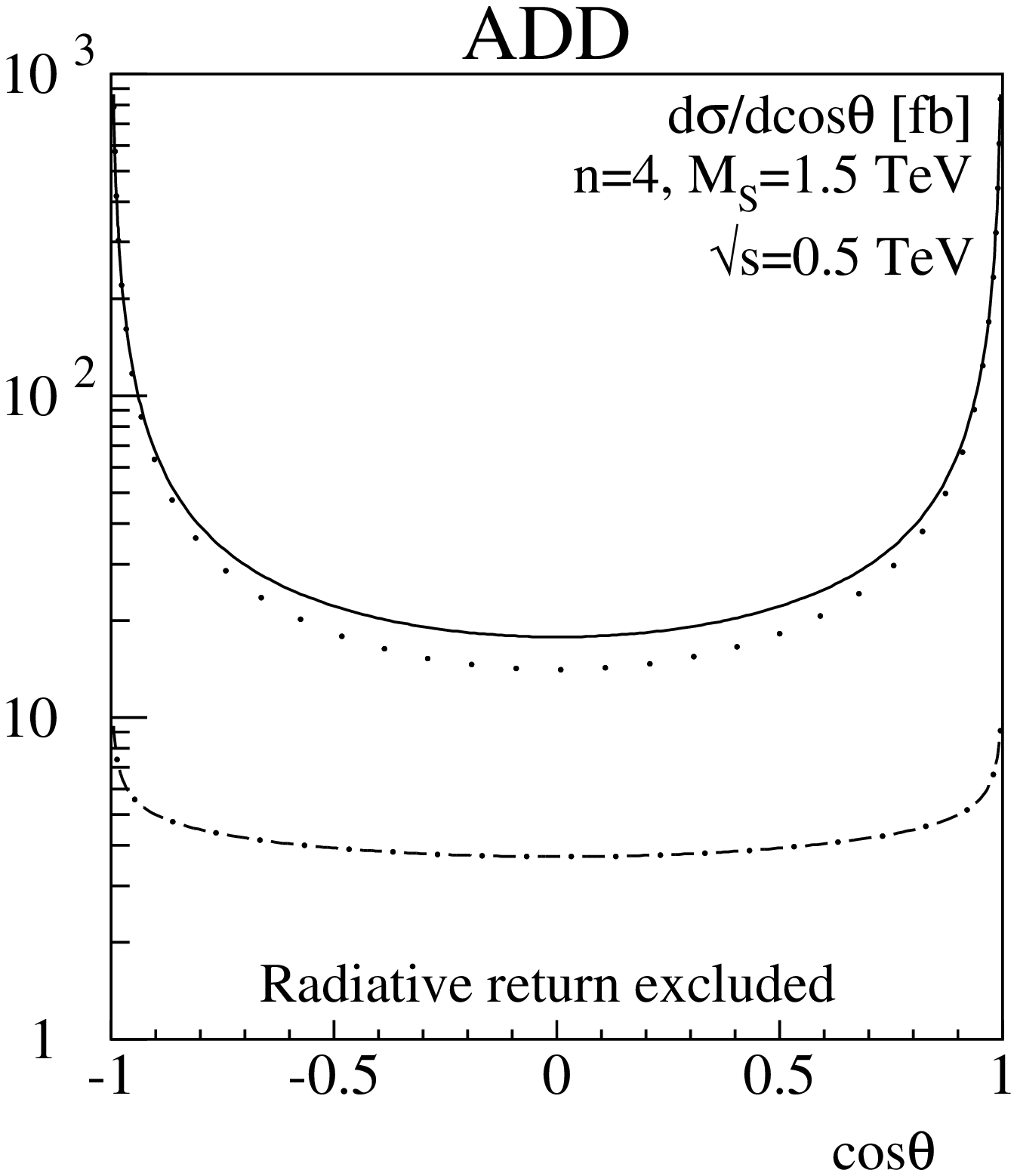}}
 \mbox{\epsfysize=7.2cm\epsffile{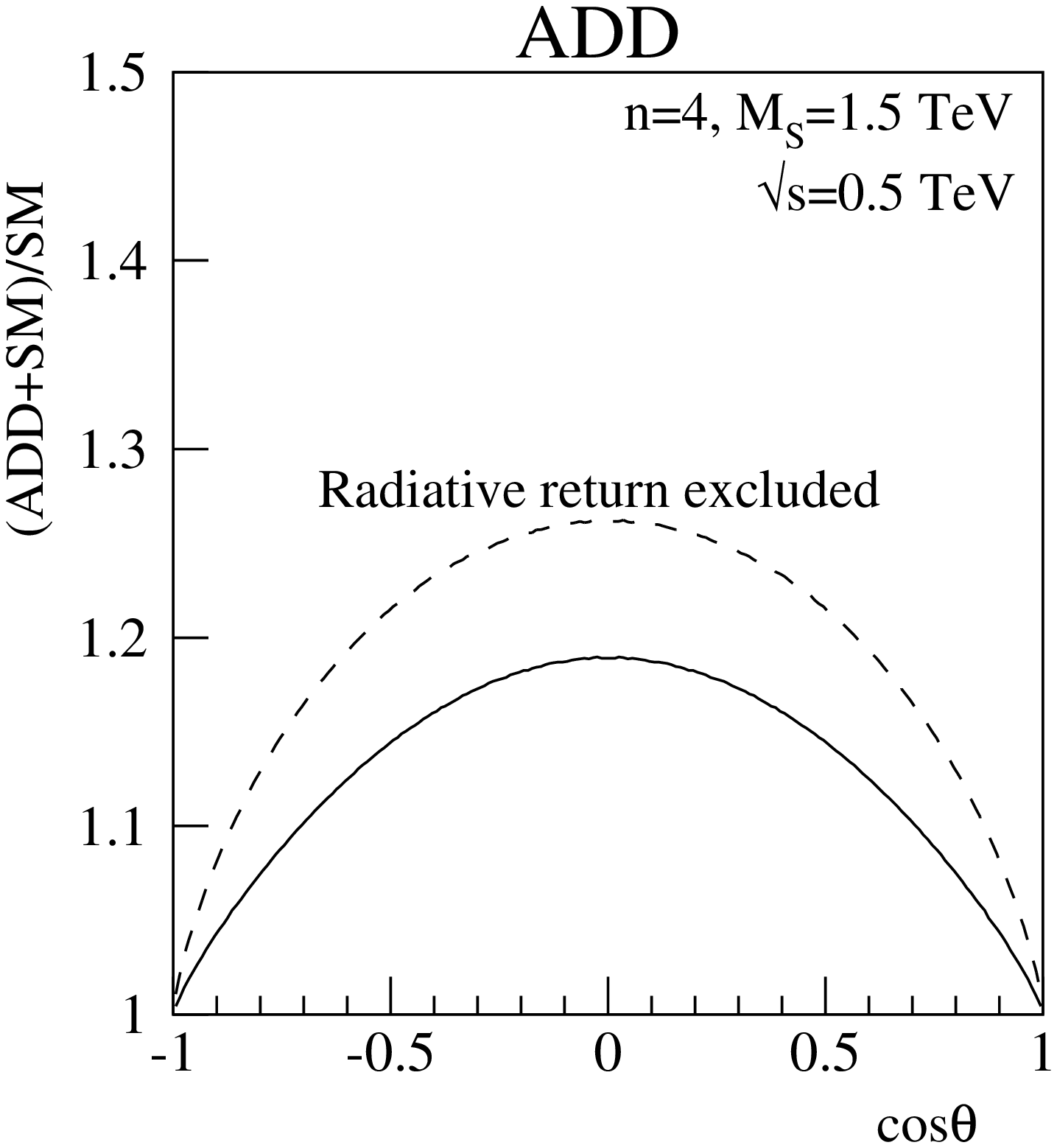}}}
\end{picture}
\vspace*{-8mm}
\caption{Photon angular distribution for $\sqrt{s}=0.5~\text{TeV}$. Left
panel: SM (dotted), contributions with graviton-exchange involved
(dash-dotted), ADD+SM (solid).  Radiative return to the $Z$ pole is excluded.
Right panel: Ratio (ADD+SM)/SM, with (solid) and without (long-dashed)
radiative return to the $Z$ pole.}
\end{center}
\end{figure}

We show in the left panel of Fig.~\ref{Fig:add-angular} the photon angular
distribution for $\sqrt{s}=0.5~\text{TeV}$, $M_S=1.5~\text{TeV}$ and $n=4$,
where radiative return to the $Z$ has been excluded. As suggested by the above
discussion, the effect of the graviton exchange is mostly to increase the
distribution in the central region, i.e., for photons making large angles with
the beams.

The enhancement at large angles, with respect to the SM, is more clearly seen
in the right panel of Fig.~\ref{Fig:add-angular}, where we show the ratio,
(ADD+SM)/SM, with and without radiative return to the $Z$. For the parameters
chosen, there is for photons perpendicular to the beam, and for the considered
parameters, an enhancement of about 25\%.


\section{The RS scenario}  \label{sec:RS}
The phenomenology of the RS scenario \cite{Randall:1999ee} differs from that
of the ADD scenario in several respects. This scenario has two 3-branes
separated in the fifth dimension, and a non-factorisable geometry, which means
that the four-dimensional metric depends on the coordinate in the fifth
dimension. It gives rise to a tower of massive KK gravitons with the mass of
the $n$'th resonance related to that of the first one, $m_1$, in the following
way \cite{Davoudiasl:1999jd}
\begin{equation} 
m_n = \frac{x_n}{x_1} m_1,
\end{equation} 
where $x_n$ are zeros of the Bessel function $J_1(x_n)=0$, with $x_1\simeq
3.83$ (not to be confused with the energy fraction carried by the $\mu^-$,
also denoted $x_1$). Therefore the mass splittings in the RS model are
non-equidistant. The mass of the first resonance is assumed to be of the order
of TeV, so only a few resonances are within reach of collider experiments. In
Fig.~\ref{Fig:rs-masses} we show the lowest states for a range of $m_1$
values.  Since there are only a few graviton resonances kinematically
available, the summation over them is straightforward.
\begin{figure}[htb]
\refstepcounter{figure}
\label{Fig:rs-masses}
\addtocounter{figure}{-1}
\begin{center}
\setlength{\unitlength}{1cm}
\begin{picture}(14,5.5)
\put(2.8,0.0)
{\mbox{\epsfysize=6cm\epsffile{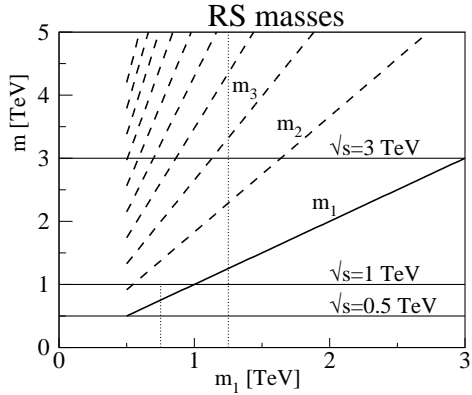}}}
\end{picture}
\vspace*{-8mm}
\caption{The lowest masses $m_i$ vs.\ $m_1$, for the RS scenario.
The vertical (dotted) lines correspond to values of $m_1$ considered
in Figs.~\ref{Fig:rs-kperp-tesla} and \ref{Fig:rs-kperp-clic}.
The horizontal lines correspond to c.m.\ energies considered.}
\end{center}
\end{figure}

The RS scenario can for our purposes be parametrized by two parameters, the
mass of the lowest massive graviton, $m_1$, and $k/\overline M_\text{Pl}$, a
dimensionless quantity typically taken in the range 0.01--0.1, effectively
giving the strength of the graviton coupling \cite{Davoudiasl:1999jd}.  The
parameter $k$ here refers to the curvature of the five-dimensional space, and
should not be confused with the photon momentum, also denoted $k$.

Expressed in terms of RS parameters, the graviton coupling, $\kappa$, of
Eqs.~(\ref{Eq:dsigma-G}) and (\ref{Eq:dsigma-G-SM}) becomes
\begin{equation} \label{Eq:kappa-RS}
\kappa = \sqrt{2} \frac{x_1}{m_1} \frac{k}{\overline M_\text{Pl}}, \qquad
\overline M_\text{Pl} = \frac{M_\text{Pl}}{\sqrt{8 \pi}} 
= 2.4 \times 10^{18}~\text{GeV},
\end{equation} 
and the widths of the resonances are given by (see \cite{Han:1999sg,
Allanach:2000nr,Dvergsnes:2002nc})
\begin{equation} 
\Gamma_n = \frac{\gamma_G}{10 \pi} x_n^2 m_n 
\left(\frac{k}{\overline M_\text{Pl}}\right)^2,
\end{equation} 
where $\gamma_G = 295/96$ (for coupling to the SM particles only).

While an RS graviton couples like an ADD graviton (apart from the strength),
the over-all phenomenology is rather different. For the two-body final states,
the RS gravitons, since they are very narrow, only contribute to the cross
section if the c.m.\ energy coincides with a graviton mass. This restriction
is lifted for the three-body final states considered here, since the diagrams
of Fig.~\ref{Fig:ee-Feynman-in} (for ISR) may resonate when $s_3$ has a
suitable value (see Eq.~(\ref{Eq:dsigma-G})), i.e., radiative return may lead
to an enhancement of the cross section.

We shall below discuss total cross sections and photon perpendicular-momentum
distributions.  The angular distributions will not be displayed for the RS
case, they are very similar to the distributions shown for the ADD case. If
$\sqrt{s}\simeq m_i$, graviton exchange will dominate, which results in a
distribution like the dash-dotted one in Fig.~\ref{Fig:add-angular}. If we are
far away from any direct resonance, the distribution will be a mixture of the
SM and graviton distributions like in the ADD case.

\subsection{Total cross sections}
In Figs.~\ref{Fig:rs-sigtot-tesla} and~\ref{Fig:rs-sigtot-clic}, we present
the total cross sections for the Bremsstrahlung process (\ref{Eq:ee-mumuga})
at four different collider energies, $\sqrt{s}=0.5$, 1, 3 and 5~TeV, as
functions of $m_1$, and for different values of $k/\overline
M_\text{Pl}=0.01$, 0.05 and 0.1.

\begin{figure}[htb]
\refstepcounter{figure}
\label{Fig:rs-sigtot-tesla}
\addtocounter{figure}{-1}
\begin{center}
\setlength{\unitlength}{1cm}
\begin{picture}(14,6.5)
\put(0,0.0)
{\mbox{\epsfysize=7cm\epsffile{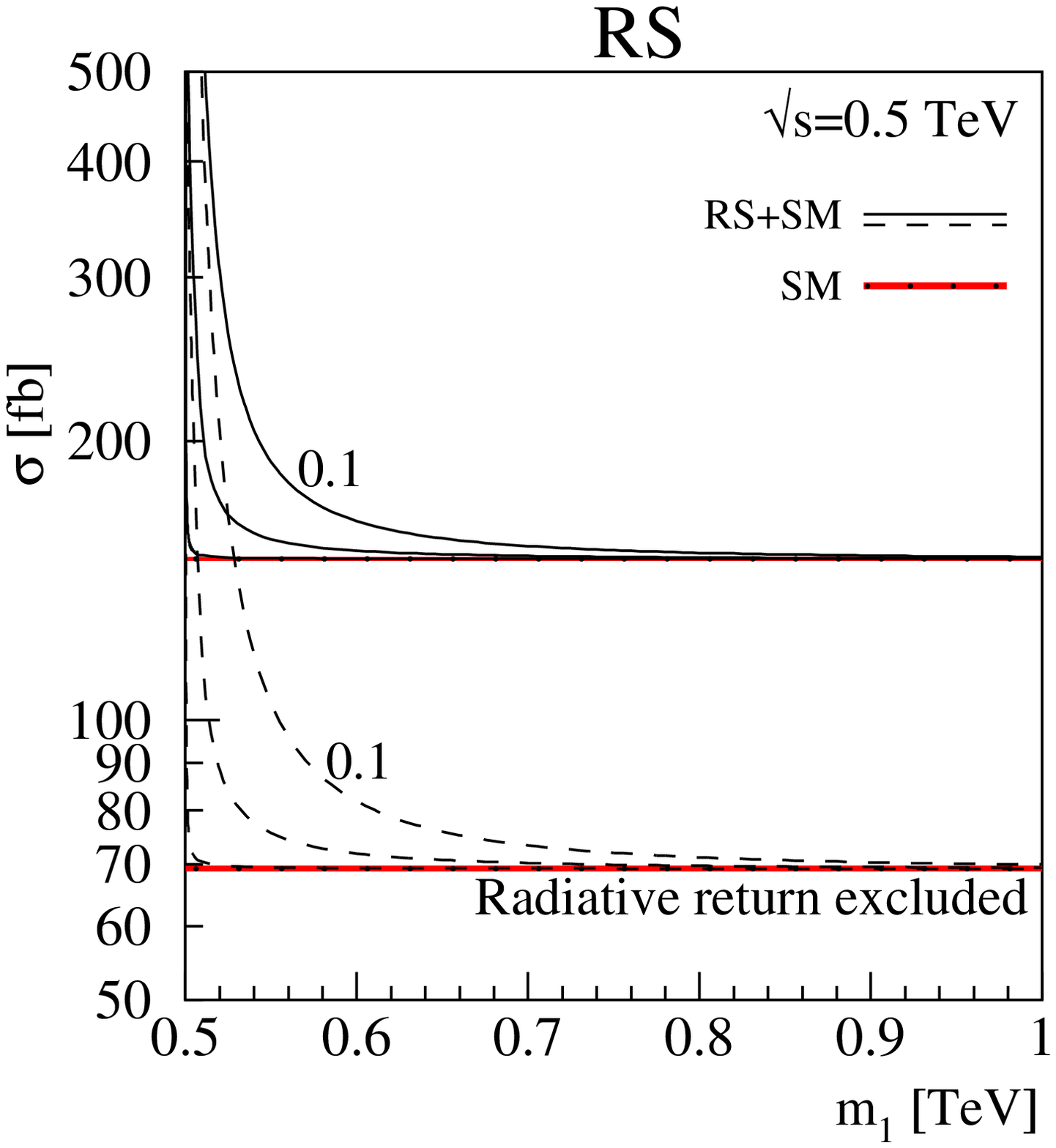}}
 \mbox{\epsfysize=7cm\epsffile{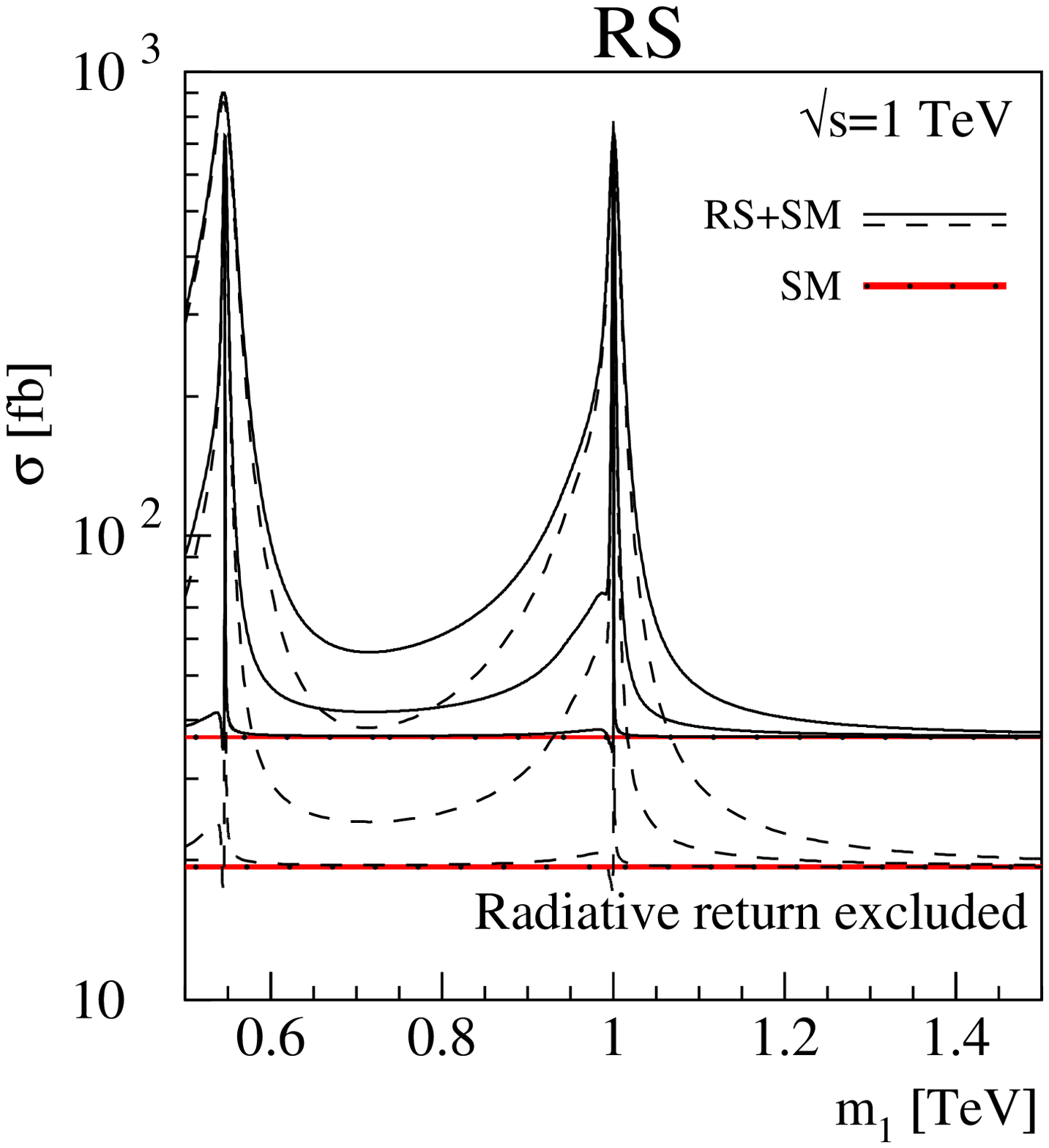}}}
\end{picture}
\vspace*{-8mm}
\caption{Total cross sections for $e^+e^-\to\mu^+\mu^-\gamma$ vs.\ $M_S$, for
$\sqrt{s}=0.5$ and 1~TeV, with (solid) and without (dashed) radiative return
to the $Z$ pole.  Three values of $k/\overline M_\text{Pl}$ are considered for
each energy; from top and down: 0.1, 0.05 and 0.01. The SM contribution is
represented by a band corresponding to $300~\text{fb}^{-1}$.}
\end{center}
\end{figure}
\begin{figure}[htb]
\refstepcounter{figure}
\label{Fig:rs-sigtot-clic}
\addtocounter{figure}{-1}
\begin{center}
\setlength{\unitlength}{1cm}
\begin{picture}(14,6.5)
\put(0,0.0)
{\mbox{\epsfysize=7cm\epsffile{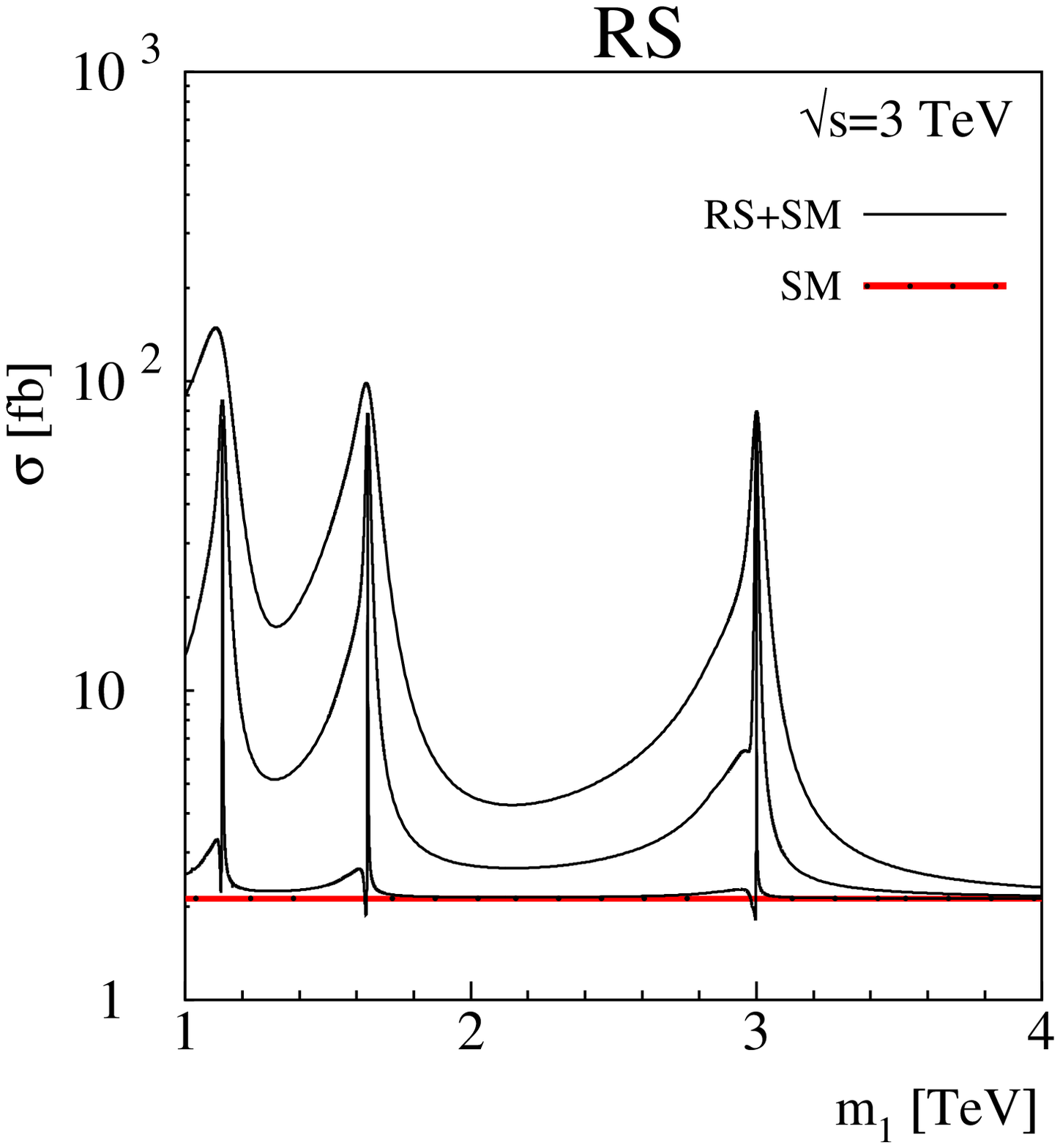}}
 \mbox{\epsfysize=7cm\epsffile{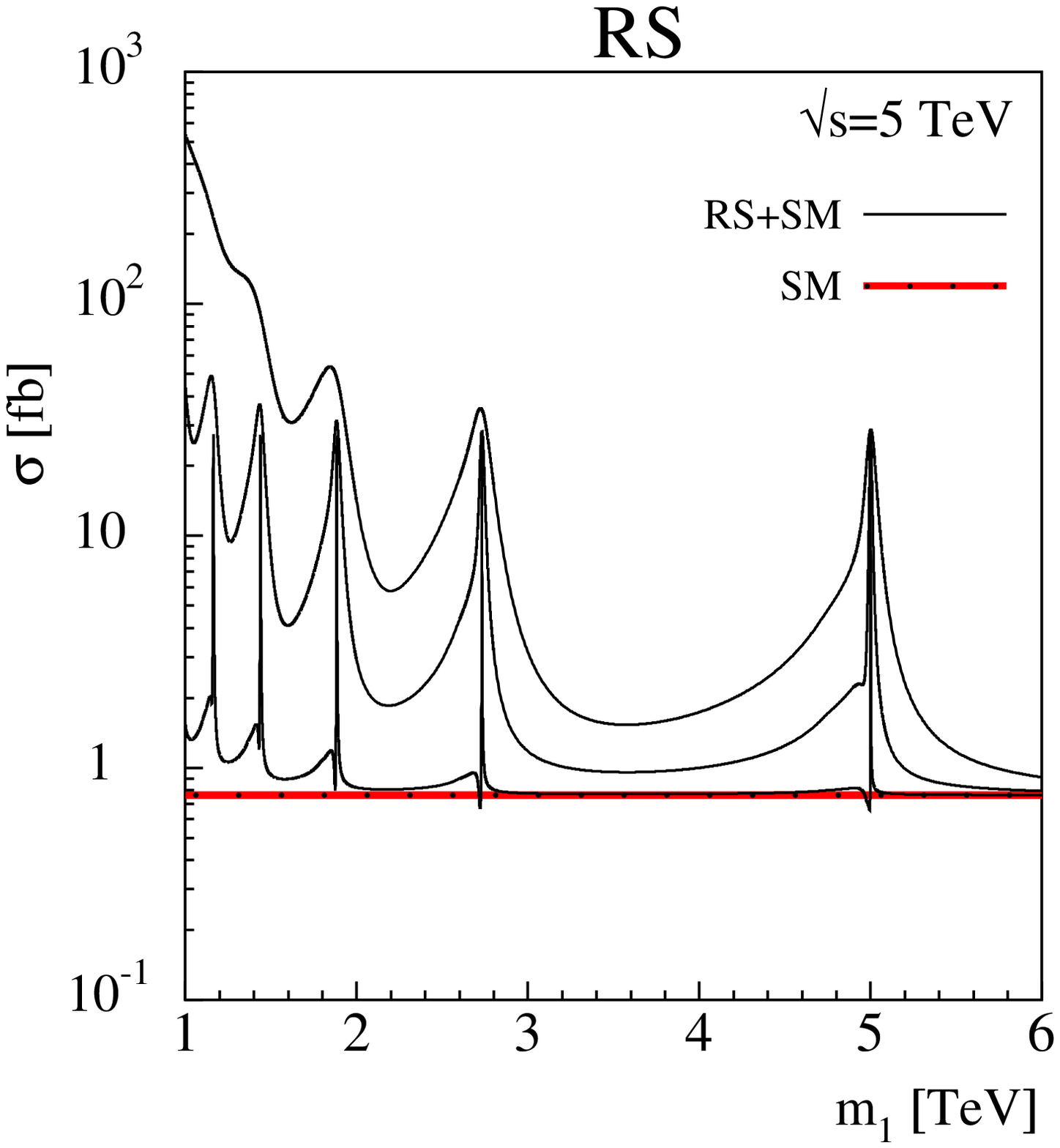}}}
\end{picture}
\vspace*{-8mm}
\caption{Total cross sections for $e^+e^-\to\mu^+\mu^-\gamma$ vs.\ $M_S$, for
$\sqrt{s}=3$ and 5~TeV.  Three values of $k/\overline M_\text{Pl}$ are
considered, like in Fig.~\ref{Fig:rs-sigtot-tesla}. The SM contribution is
represented by a band corresponding to $1000~\text{fb}^{-1}$.}
\end{center}
\end{figure}

Some of these figures have a lot of structure. Anticipating that values of
$m_1$ below the lowest considered c.m.\ accelerator energy will already be
excluded, we show in Fig.~\ref{Fig:rs-sigtot-tesla} for
$\sqrt{s}=0.5~\text{TeV}$ (left panel) only values of $m_1$ such that
$m_1>\sqrt{s}$. However, if the resonance is reasonably broad (high
$k/\overline M_\text{Pl}$), there can be a considerable increase of the cross
section for some range of $m_1$ values well above the c.m.\ energy. Like for
the ADD case, exclusion of radiative return to the $Z$ leads to an improvement
of the signal.

At the next higher energy studied, $\sqrt{s}=1~\text{TeV}$
(Fig.~\ref{Fig:rs-sigtot-tesla}, right panel), we consider a range of $m_1$
values, below the c.m.\ energy, as well as above. Apart from the obvious
resonance peak when $m_1\simeq\sqrt{s}$, there is also a sharp peak for values
of $m_1$ around 0.55~TeV. From Fig.~\ref{Fig:rs-masses} we see that this
corresponds to the second graviton, with mass $m_2$, being produced
resonantly.  We shall refer to both these cases as ``direct'' resonances,
since $\sqrt{s}=m_i$ for some $i$.

In Fig.~\ref{Fig:rs-sigtot-clic}, this phenomenon of producing higher
resonances is demonstrated for the c.m.\ energies of 3 and 5~TeV.  In the
right panel of Fig.~\ref{Fig:rs-sigtot-clic}, for $\sqrt{s}=5~\text{TeV}$, we
see for $m_1\simeq1~\text{TeV}$ and large $k/\overline M_\text{Pl}$ an
enhancement of the cross section by more than two orders of magnitude. This is
in part caused by the higher resonances being close to each other (and wide),
such that several of them can interfere. Also radiative return to lower states
contributes, as discussed below.

In this same panel, we note that there is a significant enhancement of the RS
cross section in the region around $m_1=4~\text{TeV}$, which is not compatible
with any direct resonance (when $\sqrt{s}=5~\text{TeV}$). This enhancement
is more than what can be attributed to the width of the nearby resonances, it
is caused by diagrams where the $s_3$-channel may resonate, i.e., where
$\sqrt{s_3}\simeq m_1$ and the remaining energy is carried by the photon.
\subsection{Photon perpendicular momentum distributions}

In the photon perpendicular-momentum distribution, we expect a harder spectrum
than in the SM case, as was the case for the ADD scenario. Furthermore,
resonant production of either the lowest ($m_1$) or a higher resonance ($m_i$)
can lead to a sharp edge for
\begin{equation} \label{Eq:rs-kperp-peak}
k_\perp\lsim\frac{s-m_i^2}{2\sqrt{s}},
\end{equation}
characteristic of radiative return to a lower state, $m_i<\sqrt{s}$.

\begin{figure}[htb]
\refstepcounter{figure}
\label{Fig:rs-kperp-tesla}
\addtocounter{figure}{-1}
\begin{center}
\setlength{\unitlength}{1cm}
\begin{picture}(14,7.)
\put(-1,0.0)
{\mbox{\epsfysize=7.2cm\epsffile{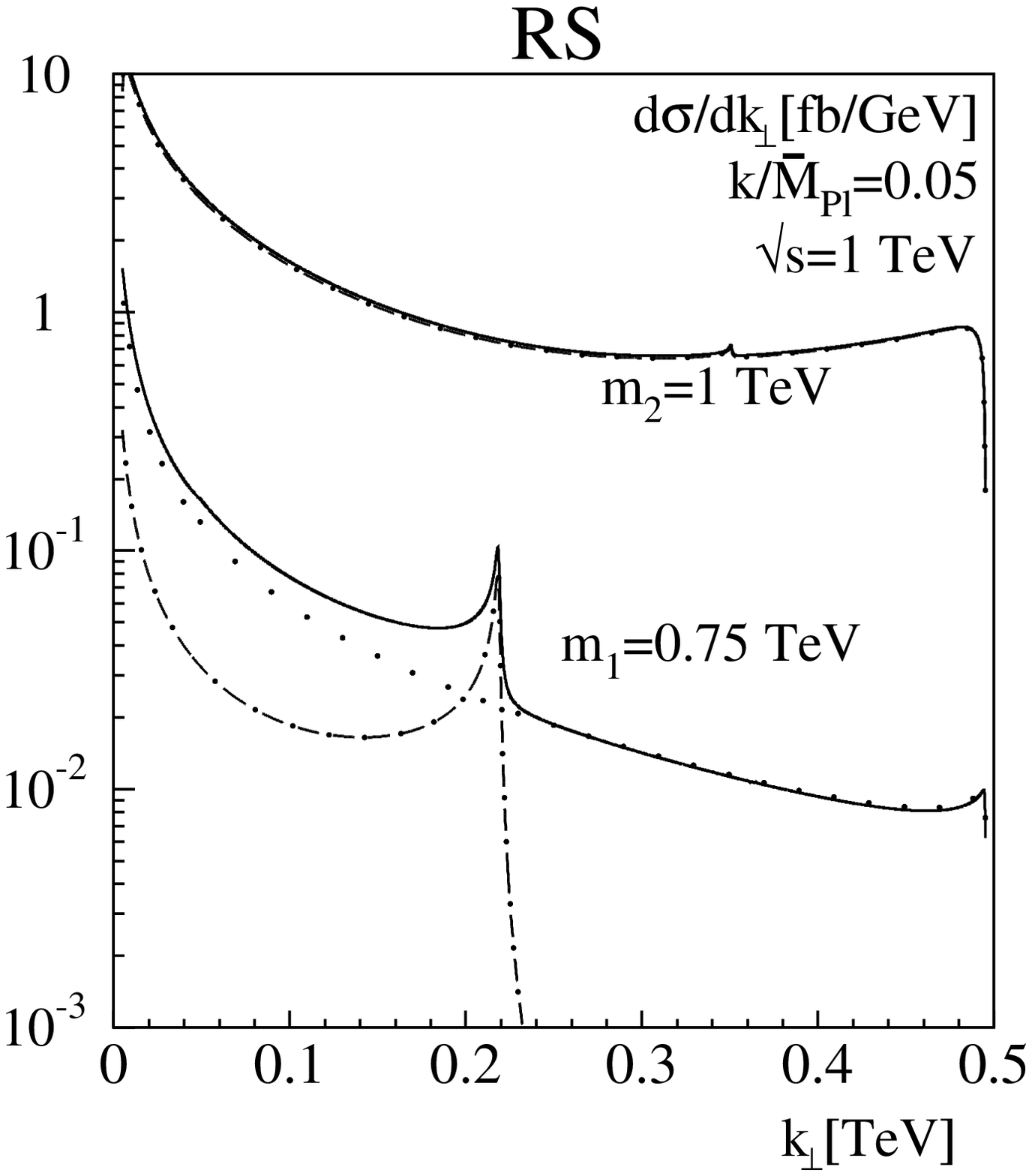}}
 \mbox{\epsfysize=7.2cm\epsffile{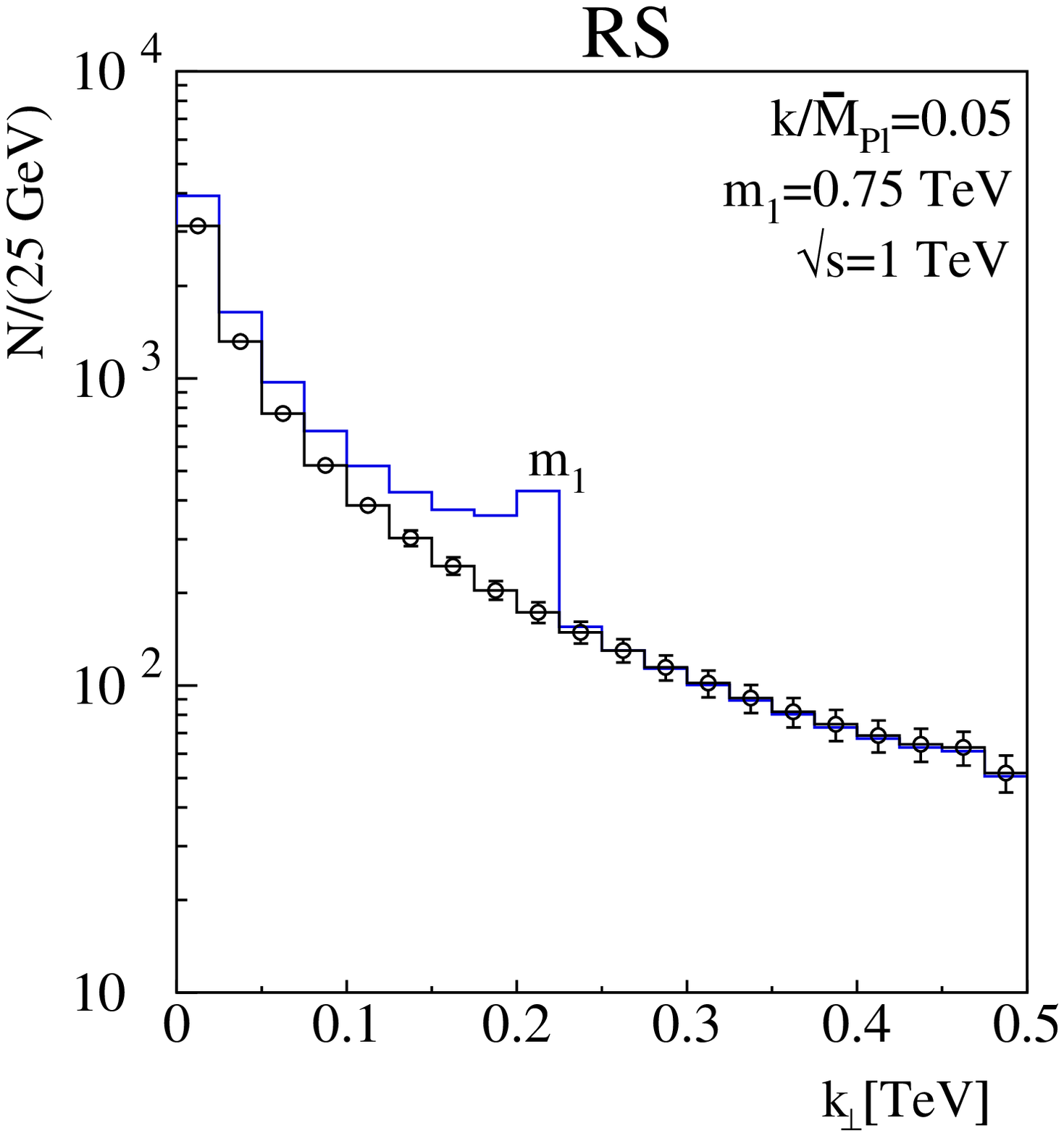}}}
\end{picture}
\vspace*{-8mm}
\caption{Photon perpendicular momentum distributions for
$\sqrt{s}=1~\text{TeV}$. Radiative return to the $Z$ is excluded.  Left panel:
Two values of $m_1$ are considered, lower curves, $m_1=0.75~\text{TeV}$, upper
curve: $m_1\simeq0.55~\text{TeV}$ chosen such that $m_2=1~\text{TeV}$.  The
graviton-related contributions are dash-dotted, the SM contribution is
dotted. Right panel: Bin-integrated $k_\perp$ distribution for
$m_1=0.75~\text{TeV}$.  The SM distribution is shown with error bars
corresponding to $\Lumint=300~\text{fb}^{-1}$.}
\end{center}
\end{figure}

\begin{figure}[htb]
\refstepcounter{figure}
\label{Fig:rs-kperp-clic}
\addtocounter{figure}{-1}
\begin{center}
\setlength{\unitlength}{1cm}
\begin{picture}(14,7.)
\put(-1,0.0)
{\mbox{\epsfysize=7.2cm\epsffile{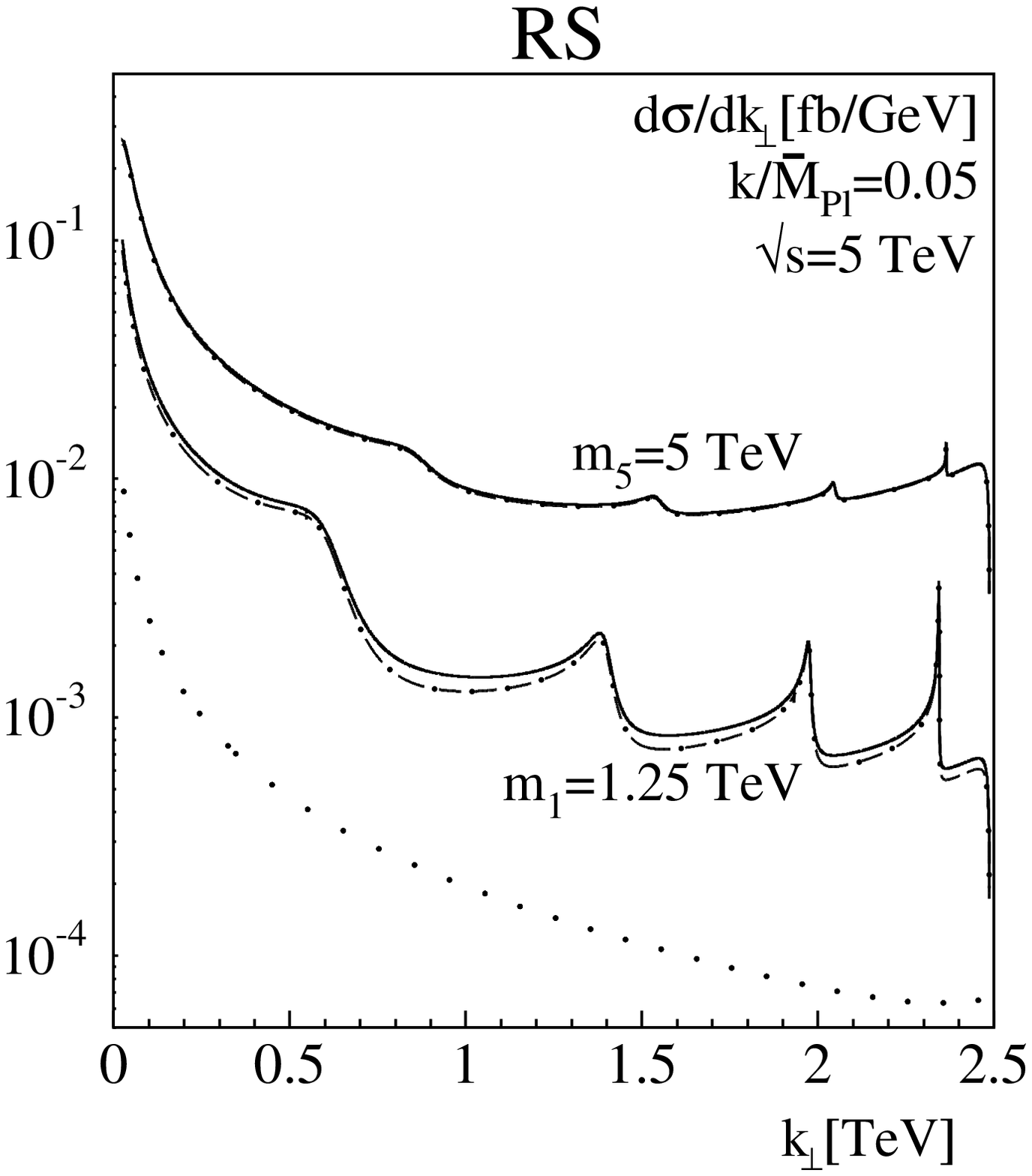}}
 \mbox{\epsfysize=7.2cm\epsffile{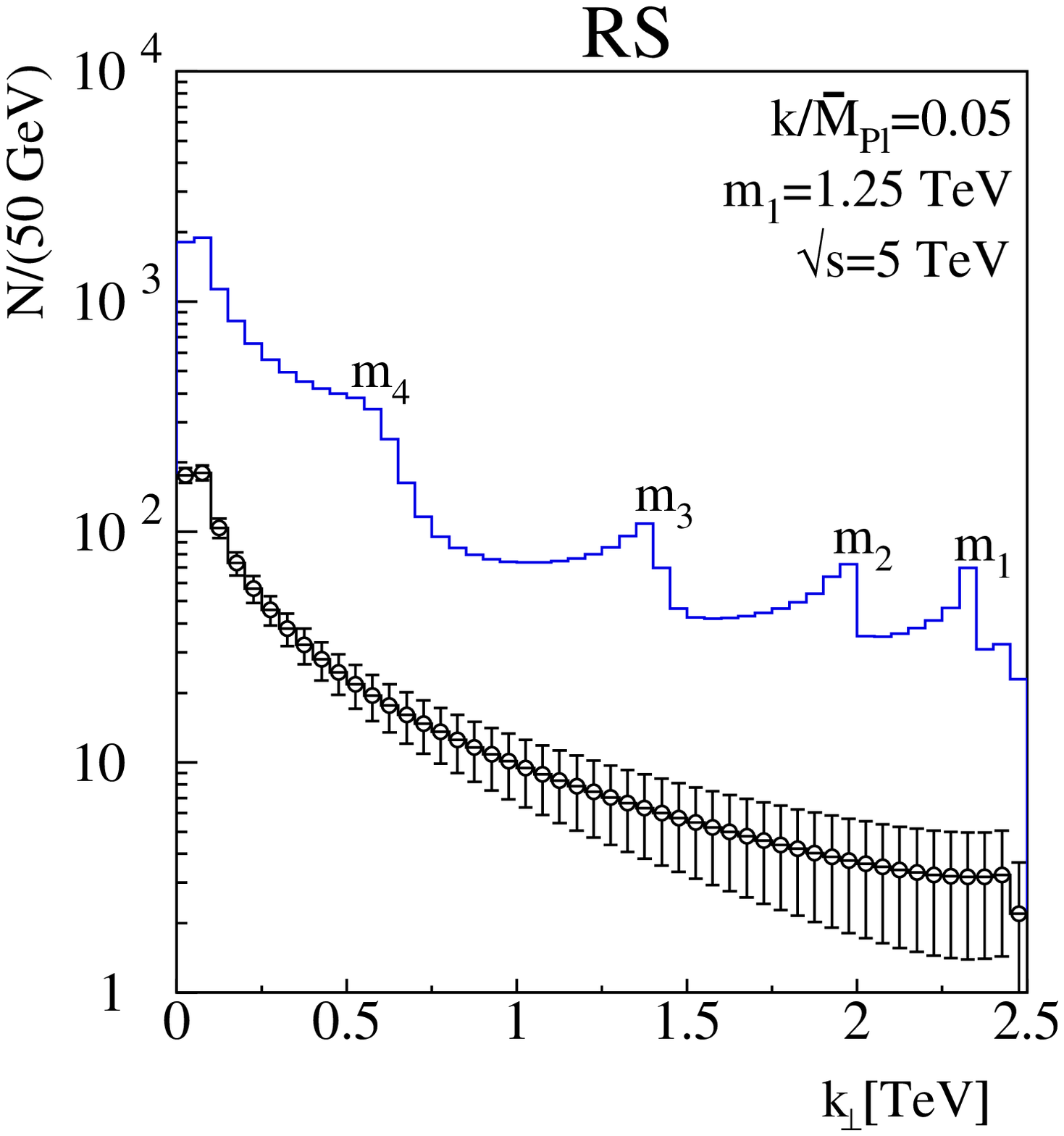}}}
\end{picture}
\vspace*{-8mm}
\caption{Photon perpendicular momentum distributions. Radiative return to the
$Z$ is excluded.  Left panel: Two values of $m_1$ are considered, lower
curves, $m_1=1.25~\text{TeV}$, upper curve: $m_1\simeq1.16~\text{TeV}$ is
chosen such that $m_5=5~\text{TeV}$.  The graviton-related contributions are
dash-dotted, the SM contribution is dotted. Right panel: Bin-integrated
$k_\perp$ distribution for $m_1=1.25~\text{TeV}$.  The SM distribution is
shown with error bars corresponding to $\Lumint=1000~\text{fb}^{-1}$.}
\end{center}
\end{figure}

Fig.~\ref{Fig:rs-kperp-tesla} is devoted to $k_\perp$ distributions for
$\sqrt{s}=1~\text{TeV}$, two values of $m_1$, and $k/\overline
M_\text{Pl}=0.05$.  The higher curves in the left panel show $k_\perp$
distributions for a reasonably low value of $m_1$, chosen such that the second
resonance coincides with the c.m.\ energy.  The distribution is for all
$k_\perp$ higher than that of the SM by more than one order of magnitude, the
excess increasing with $k_\perp$. The small structure at
$k_\perp\sim0.35~\text{TeV}$ is due to radiative return to the lower resonance
at $m_1\simeq0.55~\text{TeV}$, with the ``resonant'' $k_\perp$ given by
Eq.~(\ref{Eq:rs-kperp-peak}).  The lower curves in the left panel correspond
to a value of $m_1=0.75~\text{TeV}$ for which there is no direct
resonance. Hence, the indirect effect of radiative return becomes more
visible, there is a distinct enhancement at the value of $k_\perp$
corresponding to (\ref{Eq:rs-kperp-peak}).

In the right panel we show the binned distribution for $m_1=0.75~\text{TeV}$
together with the SM prediction with error bars corresponding to
$\Lumint=300~\text{fb}^{-1}$.  The bin width has been chosen as 25~GeV
\cite{Aguilar-Saavedra:2001rg}.  The enhancement related to radiative return
to the $m_1$ is clearly visible above the statistical noise.

In Fig.~\ref{Fig:rs-kperp-clic} we show $k_\perp$ distributions for
$\sqrt{s}=5~\text{TeV}$ and two values of $m_1$.  The upper curves in the left
panel correspond to a value of $m_1$ for which there is a direct resonance
corresponding to $m_5=\sqrt{s}$. The spectrum is very hard, and small features
corresponding to radiative return to all the lower resonances are seen.  The
middle curves, which are about an order of magnitude above the SM background
(dotted), correspond to a value $m_1=1.25~\text{TeV}$ for which there is no
direct resonance. As can be seen from Fig.~\ref{Fig:rs-masses}, $m_1$, $m_2$,
$m_3$ and $m_4$ are accessible, and show up as peaks in the $k_\perp$
distribution.  In the right panel we show the binned distribution for
$m_1=1.25~\text{TeV}$ together with the SM prediction with error bars
corresponding to $\Lumint=1000~\text{fb}^{-1}$.  The enhancements related to
radiative return to $m_1, \ldots, m_4$ are clearly visible above the
statistical noise.  Another distinctive feature is that the interference
between different gravitons leads to a significant enhancement of the cross
section over the SM background for all values of $k_\perp$.
\section{Summary}  \label{sec:concl}

While the three-body cross section is lower than those of the corresponding
two-body final states $\mu^+\mu^-$ and $\gamma\gamma$ by a factor of order
$\alpha/\pi$, and therefore is unlikely to be a discovery channel for
massive-graviton effects, it has some distinctive features which differ from
the SM and may help distinguishing between the different scenarios.  First of
all, the $k_\perp$ distribution is harder than in the SM. This applies to both
the ADD and RS scenarios, and can be particularly important in the RS
scenario, if the graviton has a moderately strong coupling (determined by
$k/\overline M_\text{Pl}$). Also, the photon angular distribution can have a
significant enhancement at large angles.

In the ADD scenario, where the $k_\perp$ distribution is rather smooth, of the
order of one year of running would be sufficient to see this hardening of the
photon spectrum, for values of $M_S$ up to about twice the c.m.\ energy.

In the RS scenario, ISR opens up the possibility of radiative return to the KK
graviton resonances within the kinematically accessible range. This can lead
to characteristic perpendicular-momentum distributions, and an increase in the
cross section even when the c.m.\ energy is far away from any resonance.

Radiative return to the $Z$ is also possible through ISR, but can be removed
by a cut. The statistical significance of the signal can improve significantly
when such a cut is included.

Here we have considered a final state with a lepton pair accompanied by a
photon. It would also be of interest to consider different final states like
$q \bar q \gamma$ (two jets and a photon) or even gluon Bremsstrahlung,
$e^+e^-\to q \bar q g$ (three jets) in future analyses. In the latter case,
the result would however be different from the case considered here (after the
trivial substitutions for other coupling constants and colour factors). The
reason for this difference is that the gluon can only come from the quark
line, the ISR contribution would only yield photons, and therefore be of
higher order compared to $e^+e^-\to\text{three jets}$.

\bigskip

\medskip
\leftline{\bf Acknowledgment} 
\par\noindent 
This research has been supported in part by the Research Council of Norway.

\appendix
\renewcommand{\theequation}{\thesection\arabic{equation}}

\section*{Appendix A: Angular- and energy-distribution functions}
\setcounter{equation}{0}
\renewcommand{\thesection}{A}

The angular and energy distributions of the different contributions to the
cross section are in Eqs.~(\ref{Eq:dsigma-G}), (\ref{Eq:dsigma-SM}) and
(\ref{Eq:dsigma-G-SM}) expressed in terms of the functions
$X_{AA}(x_3,\eta,\cos\theta)$ etc., where $\eta=x_1-x_2$.  It is convenient to
introduce the abbreviations:
\begin{alignat}{2}
z_{a} &= 8x_3^4 - 12x_3^2 + 12x_3 -3, &\quad
z_{j} &= 2x_3^2 + 2x_3 - 1, \nonumber\\
z_{b} &= 3(1 - 2x_3), &\quad
z_{k} &= 4x_3^2 + 4x_3 - 3, \nonumber\\
z_{c} &= 2x_3^2 - 2x_3 + 1, &\quad
z_{l} &= 4x_3^2 - 8x_3 + 3, \nonumber\\
z_{d} &= 4x_3^2 - 2x_3 +1, &\quad
z_{m} &= 4x_3^2 - 5x_3 + 3, \nonumber\\
z_{e} &= 2(1-x_3)^2, &\quad
z_{n} &= 4x_3^2 - 20x_3 + 15, \nonumber\\
z_{f} &= 4x_3^2 - 10x_3 + 5, &\quad
z_{o} &= 8x_3 - 3, \nonumber \\
z_{g} &= 2x_3^2 - 6x_3 + 3, &\quad
z_{p} &= 4x_3 - 3, \nonumber\\
z_{h} &= 4x_3^2 - 14x_3 + 7, &\quad
z_{q} &= 6x_3^2 - 7x_3 + 3, \nonumber\\
z_{i} &= 8x_3^4 - 80x_3^3 + 180x_3^2 - 140x_3 + 35, &\quad
z_{r} &= 24x_3^2 - 40x_3 + 15.
\end{alignat}

Here we give the functions defining the different contributions. We start with
pure graviton exchange [see Eq.~(\ref{Eq:dsigma-G})]:
\begin{align}
\label{Eq:XG}
X_{AA}(x_3,\eta,\cos\theta) 
&= \frac{\ta_0(x_3,\eta) + \ta_2(x_3,\eta)\cos^2\theta
        +\ta_4(x_3,\eta)\cos^4\theta}
                                 {x_3^6(1-\cos^2\theta)}, \nonumber \\
X_{AB}(x_3,\eta,\cos\theta) 
&= (1-x_3) \frac{\ta_1(x_3,\eta)\cos\theta 
                +\ta_3(x_3,\eta)\cos^3\theta}
                                 {x_3^5}, \nonumber \\
X_{BB}(x_3,\eta,\cos\theta) 
&= \frac{\ta_0(x_3,\eta) + \ta_2(x_3,\eta)\cos^2\theta
        +\ta_4(x_3,\eta)\cos^4\theta}
                                 {x_3^4(1-2x_3)(x_3^2-\eta^2)},
\end{align}
with
\begin{align}
\ta_0(x_3,\eta) &= - \eta^4 z_{a} - \eta^2 x_3^2 z_{b}z_{c} 
                 + x_3^4 z_{d}z_{e}, \nonumber \\
\ta_1(x_3,\eta) &= - 2 \eta^3 z_{b} + \eta x_3^2 z_{b}, \nonumber \\
\ta_2(x_3,\eta) &= - 2\eta^4 z_{b}z_{f} + 3\eta^2 x_3^2 z_{b}z_{g} 
                 - x_3^4 z_{b}z_{c}, \nonumber\\
\ta_3(x_3,\eta) &= 2\eta^3 z_{h} - 2 \eta x_3^2 z_{b}, \nonumber\\
\ta_4(x_3,\eta) &= \eta^4 z_{i} - 2\eta^2 x_3^2 z_{b}z_{f} - x_3^4 z_{a}.
\end{align}

Next we give the pure SM terms [see Eq.~(\ref{Eq:dsigma-SM})]:
\begin{align}
\label{Eq:XSM}
X_{C_\gamma C_\gamma}(x_3,\eta,\cos\theta) &= 
                     \frac{\tb_0(x_3,\eta) + \tb_2(x_3,\eta)\cos^2\theta}
                          {x_3^4(1-2x_3)(1-\cos^2\theta)}, \nonumber \\
X_{C_\gamma C_Z}(x_3,\eta,\cos\theta) &= 
X_{C_Z C_\gamma}(x_3,\eta,\cos\theta) = v_e v_\mu X_{C_\gamma C_\gamma}
                    +a_e a_\mu \frac{\tb_1(x_3,\eta)\cos\theta }
                          {x_3^4(1-2x_3)(1-\cos^2\theta)}, \nonumber \\
X_{C_Z C_Z}(x_3,\eta,\cos\theta) &= (a_e^2+v_e^2)(a_\mu^2+v_\mu^2)
                    X_{C_\gamma C_\gamma}
                    +4a_e a_\mu v_e v_\mu \frac{\tb_1(x_3,\eta)\cos\theta }
                          {x_3^4(1-2x_3)(1-\cos^2\theta)}, \nonumber \\
X_{C_\gamma D_\gamma}(x_3,\eta,\cos\theta) &= 
                     (1-x_3)\frac{\eta\cos\theta }
                          {x_3^3}, \nonumber \\
X_{C_\gamma D_Z}(x_3,\eta,\cos\theta) &=
X_{C_Z D_\gamma}(x_3,\eta,\cos\theta) = (1-x_3) \frac{v_e v_\mu
                     \eta \cos\theta - a_e a_\mu x_3} {x_3^3}, \nonumber \\
X_{C_Z D_Z}(x_3,\eta,\cos\theta) &= (1-x_3)\frac{(a_e^2+v_e^2)(a_\mu^2+v_\mu^2)
           \eta \cos\theta - 4a_e a_\mu v_e v_\mu x_3} {x_3^3}, \nonumber \\
X_{D_\gamma D_\gamma}(x_3,\eta,\cos\theta) &= 
                     \frac{\tb_0(x_3,\eta) + \tb_2(x_3,\eta)\cos^2\theta}
                          {x_3^2(x_3^2 - \eta^2)}, \nonumber \\
X_{D_\gamma D_Z}(x_3,\eta,\cos\theta) &= 
X_{D_Z D_\gamma}(x_3,\eta,\cos\theta) =  v_e v_\mu 
                  X_{D_\gamma D_\gamma}
                  +a_e a_\mu \frac{\tb_1(x_3,\eta)\cos\theta}
                          {x_3^2(x_3^2-\eta^2)}, \nonumber \\
X_{D_Z D_Z}(x_3,\eta,\cos\theta) &= (a_e^2+v_e^2)(a_\mu^2+v_\mu^2)
                  X_{D_\gamma D_\gamma}
                  +4a_e a_\mu v_e v_\mu \frac{\tb_1(x_3,\eta)\cos\theta}
                          {x_3^2(x_3^2-\eta^2)},
\end{align}
with
\begin{align}
\tb_0(x_3,\eta) &= \eta^2 z_{j} + x_3^2 z_{g}, \nonumber \\
\tb_1(x_3,\eta) &= -4\eta x_3 z_{c}, \nonumber \\
\tb_2(x_3,\eta) &= \eta^2 z_{g} + x_3^2 z_{j}.
\end{align}
Vector and axial couplings are normalized to 
$v_f=T_f-2Q_f\sin^2\theta_W$, $a_f=T_f$, with $T_f$ the isospin.

Then we list the graviton-SM interference terms. First we have the pure ISR
and FSR terms:
\begin{align}
\label{Eq:XAC-XBD}
X_{AC_\gamma}(x_3,\eta,\cos\theta) 
&= \frac{\tc_1(x_3,\eta)\cos\theta + \tc_3(x_3,\eta)\cos^3\theta}
                          {x_3^5(1-\cos^2\theta)}, \nonumber \\
X_{AC_Z}(x_3,\eta,\cos\theta) 
&= 2v_e v_\mu X_{AC_\gamma} 
+a_e a_\mu \frac{\tc_0(x_3,\eta)+\tc_2(x_3,\eta)\cos^2\theta}
                          {x_3^5(1-\cos^2\theta)}, \nonumber \\
X_{BD_\gamma}(x_3,\eta,\cos\theta) 
&= \frac{\tc_1(x_3,\eta)\cos\theta + \tc_3(x_3,\eta)\cos^3\theta}
                          {x_3^3(x_3^2-\eta^2)}, \nonumber \\
X_{BD_Z}(x_3,\eta,\cos\theta) 
&= 2 v_e v_\mu X_{BD_\gamma}
 +a_e a_\mu \frac{\tc_0(x_3,\eta) + \tc_2(x_3,\eta)\cos^2\theta}
                          {x_3^3(x_3^2-\eta^2)},
\end{align}
with
\begin{align}
\tc_0(x_3,\eta) &= -3\eta^2 x_3 z_{c} + x_3^3 z_{c}, \nonumber \\
\tc_1(x_3,\eta) &= \eta^3 z_{b} - x_3^2 \eta z_{b}, \nonumber \\
\tc_2(x_3,\eta) &= 9\eta^2 x_3 z_{c} - 3x_3^3  z_{c}, \nonumber \\
\tc_3(x_3,\eta) &= -\eta^3 z_{f} + x_3^2 \eta z_{b}.
\end{align}

Finally we have the graviton-SM interference terms where one diagram is ISR
and the other one is FSR. The terms with graviton exchange in the ISR diagram
are:

\begin{align}
\label{Eq:XAD}
X_{AD_\gamma}(x_3,\eta,\cos\theta) &= (1-2x_3) 
                     \frac{\td_0(x_3,\eta) + \td_2(x_3,\eta)\cos^2\theta}
                          {x_3^4}, \nonumber \\
X_{AD_Z}(x_3,\eta,\cos\theta) &= v_e v_\mu X_{AD_\gamma}
                    +a_e a_\mu (1-2x_3) 
                    \frac{\td_1(x_3,\eta) \cos\theta}
                          {x_3^4},
\end{align}
with
\begin{align}
\td_0(x_3,\eta) &= -\eta^2 z_{k} - x_3^2 z_{l}, \nonumber \\
\td_1(x_3,\eta) &= 4\eta x_3 z_{m}, \nonumber \\
\td_2(x_3,\eta) &= -\eta^2 z_{n} - x_3^2 z_{k}.
\end{align}
The terms with graviton exchange in the FSR diagram are:
\begin{align}
\label{Eq:XBC}
X_{BC_\gamma}(x_3,\eta,\cos\theta) &= 
                   \frac{\te_0(x_3,\eta) + \te_2(x_3,\eta)\cos^2\theta}
                          {x_3^4(1-2x_3)}, \nonumber \\
X_{BC_Z}(x_3,\eta,\cos\theta) &= v_e v_\mu X_{BC_\gamma}
          +a_e a_\mu \frac{\te_1(x_3,\eta)\cos\theta}
               {x_3^4(1-2x_3)},
\end{align}
with
\begin{align}
\te_0(x_3,\eta) &= -\eta^2 z_{o} + x_3^2 z_{p}, \nonumber \\
\te_1(x_3,\eta) &= 4\eta x_3 z_{q}, \nonumber \\
\te_2(x_3,\eta) &= -\eta^2 z_{r} - x_3^2 z_{o}.
\end{align}


\goodbreak

\clearpage
\end{document}